\documentstyle[preprint,prb,aps]{revtex}

\newcommand{\bk}{{\bf k}}
\newcommand{\bq}{{\bf q}}
\newcommand{\br}{{\bf r}}
\newcommand{\bR}{{\bf R}}
\newcommand{\bG}{{\bf G}}
\newcommand{\phif}{{\varphi_6^\infty}}
\newcommand{\dn}{{\delta n}}

\begin{document}
\draft

\title{Vortex Lattice Melting in 2D Superconducting Networks and Films}
\author{M. Franz and S. Teitel}
\address{Department of Physics and Astronomy, University of Rochester,
   Rochester, NY 14627}
\date{\today}
\maketitle

\begin{abstract}
We carry out an extensive Monte Carlo study of phase transitions
in 2D superconducting networks, in an applied magnetic field,
for square and honeycomb geometries.
We consider both systems with
a dilute vortex density $1/q$, and
dense systems near ``full frustration" with vortex density $1/2-1/q$.
The dilute case gives the continuum limit as
$q\rightarrow \infty$, and serves as a
model for a uniform superconducting film.
For this dilute  case, we find a
transition temperature $T_c\sim 1/q$, at which the vortex lattice unpins from
the network and forms a ``floating solid'' phase.
At a higher temperature $T_m$, this floating solid melts into a
vortex liquid.  We analyze the transition at $T_m$ according
to the Kosterlitz-Thouless theory of dislocation mediated melting in 2D.
While we find a discontinuous jump in the vortex shear modulus at $T_m$
which is consistent with this theory, we find (in opposition to this theory)
that the transition is weakly first order, and we find no evidence for a
hexatic liquid phase.
For the case near full frustration, we find that the system can be described
in terms of the density of defects in an otherwise fully frustrated vortex
pattern.  These dilute defects result in similar behavior as that found in the
dilute vortex system, with pinned, floating, and liquid defect phases.

\end{abstract}


%#############################################################################

\newpage
\narrowtext

\section{Introduction}

Two dimensional (2D) periodic superconducting networks,
and in particular arrays of Josephson junctions, have served as a
convenient theoretical and experimental model system in terms of which one
can study, in a well controlled way, the effects of thermal fluctuations
and pinning, on vortex structures and phase coherence in 2D
superconductors \cite{nato}.
Such 2D superconductors have received renewed attention recently
with the observation that many of the high temperature superconductors consist
of weakly coupled layers, and so for some range of parameters may display
effectively two dimensional behavior \cite{ffh,gk}.
One focus of this renewed interest has been concerned with the
melting of the 2D vortex lattice, induced by an applied magnetic field,
in a uniform continuous superconducting film.  Controversy has resulted
as to whether such a vortex lattice
even exists at any finite temperature, or whether a vortex liquid is
the only thermodynamically stable state \cite{moore}.
In this work we address the thermodynamic behavior of vortex structures
in 2D superconducting systems.  Our focus will be on behavior in discrete
periodic networks, however our results will also yield conclusions concerning
the behavior of uniform films.

Despite a decade of theoretical work, many fundamental questions remain
unresolved concerning the nature of the phase transitions in 2D
superconducting networks.
When a uniform transverse magnetic field is applied, it induces a fixed
density of vortices into the network, as in the mixed state of a type II
superconductor.  However, unlike a uniform superconductor, for which the ground
state is a periodic triangular lattice of equally spaced vortices, the discrete
network structure serves as an effective periodic pinning potential, which at
low temperatures confines the vortices to sit at the centers of the unit cells
of the network.  This can result in novel vortex structures at low temperature,
determined by the competition between the repulsive vortex-vortex interaction,
and the periodic pinning potential induced by the
network \cite{tj,lobb,straley}.
Finding the ground state vortex structure for an arbitrary value of vortex
density,
for a given periodic network, remains an unsolved problem.
The phase transitions at finite temperature have remained largely
unexplored except for a few of the simplest
cases\cite{kosterlitz,half,jrl1,lee}.

An early conjecture by Teitel and Jayaprakash\cite{tj} (TJ) argued that
the superconducting
transition in such networks would be governed by commensurability effects.
If one measures the dimensionless vortex density $f$ as the number
of magnetic field induced vortices per unit cell of the network, they
predicted that for rational $f=p/q$, the transition temperature
would vary discontinuously as $T_c(p/q)\sim 1/q$.  While experimental
evidence for high order commensurability effects
has been reported in Josephson junction arrays, \cite{martinoli}
simulations by Halsey\cite{halsey} have
challenged this conjecture for large $q$.  A similar conjecture by
TJ\cite{tj}
concerning the behavior of the ground state critical current, $i_c(f)$,
has since been disproven in simulations by Lobb and co-workers,
\cite{barrier,expt1}
and by Straley, \cite{straley2}
who argue that as $f$ varies, $i_c(f)$ has a lower non-zero limit
determined by the single body effects of a non-interacting vortex in a
periodic pinning potential; this conclusion has also been arrived
at analytically by Vallat and Beck \cite{beck}.
However the validity of the TJ conjecture with respect to $T_c(f)$,
which is intrinsically determined by many body effects,
has remained unresolved.

In this paper we attempt to study the TJ conjecture systematically,
by carrying out Monte Carlo simulations of superconducting networks
for two special classes of vortex density.
We first consider the dilute case of vortex densities $f=1/q$, $q$ integer,
for both square and honeycomb networks.
This dilute case, as $q\to\infty$, can equivalently be viewed as the
continuum limit, in which the lattice spacing of the periodic network
decreases to zero for a fixed areal density of vortices.  Our results for this
case therefore also address the problem of vortex lattice melting in a uniform
continuous superconducting film.
Secondly, we consider vortex
densities $f=1/2-1/q$, close to full frustration, on the square network.

Our results may be summarized as follows.
For the dilute case with large $q$, the low temperature state is a
Bravais lattice of vortices, with long range translational order,
pinned commensurably to the periodic network.
At a critical temperature $T_c(1/q)\sim 1/q$, there is a sharp first
order phase transition to a floating triangular vortex lattice, which
is depinned from the periodic potential of the network.  This
floating lattice displays the algebraic translational correlations
characteristic of a 2D vortex lattice in a uniform continuum.  The
depinning transition $T_c(f)$ satisfies the TJ conjecture, and
marks the loss of true d.c. superconductivity in the network, due to the
flux flow resistance which will result from drift of the unpinned
vortex lattice.  At a higher $T_m$, which becomes independent of $q$
as $1/q\to 0$, this floating vortex lattice melts into an isotropic
vortex liquid.  We analyze this transition according to the
theory of dislocation mediated
melting in 2D, due to Kosterlitz and Thouless, \cite{ktmelt}
Nelson and Halperin, \cite{nhmelt} and Young\cite{young}
(KTNHY).  While we find good agreement with certain predictions of
this KTNHY theory, we find evidence that the second order melting transition
predicted by KTNHY is pre-empted by a weak first order transition.

For the close to fully frustrated case, $f=1/2-1/q$,
the ground state is everywhere like that of $f=1/2$ (a checkerboard
pattern of vortices on alternating sites), except for a superimposed
commensurate Bravais lattice of missing vortices, or ``defects,"
so as to give the desired density $f<1/2$.
The transitions in this system are then governed by the behavior of these
defects.  Upon heating, there is first a
depinning transition $T_c(f)$ of the defect Bravais lattice into
a floating triangular defect lattice; this depinning
follows the TJ conjecture, $T_c(f)\sim 1/q$,
and marks the loss of true d.c. superconductivity.  At a higher $T_m$, the
floating defect lattice melts into an isotropic defect liquid.  Finally,
at a higher $T_{m^\prime}$, there is an additional sharp transition
representing the disordering of the vortices forming the $f=1/2$ like
background.

The remainder of our paper is organized as follows.  In Section II
we present the theoretical model used to describe the superconducting network,
and its relation to a uniform superconducting film.  We
review the KTNHY theory of 2D melting, and discuss the observables we
measure and the methods we use to analyze our data.  Finally we
describe our Monte Carlo procedure.
In Section III we present our results for the dilute case $f=1/q$ on
a honeycomb network.  This corresponds to vortices on the dual
triangular lattice of sites.  We use finite size scaling to
test in detail the predictions of KTNHY.  In Section IV we present
our results for the dilute
case $f=1/q$ on a square network.  In Section V we present our
results for the dense
case of $f=5/11$ on a square lattice, and infer the behavior for more
general densities $f=1/2-1/q$.  In Section VI we present our discussion
and conclusions.

%%%%%%%%%%%%%%%%%%%%%%%%%%%%%%%%%%%%%%%%%%%%%%%%%%%%%%%%%%%%%%%%%%%%%%%%%%%%%%
\section{Model and Methods of Analysis}

\subsection{Model for a superconducting network}

A two dimensional superconducting network in a magnetic field, is
described by the Hamiltonian,
\begin{equation}
{\cal H}[\theta_i]=\sum_{\langle ij\rangle}U(\theta_i-\theta_j-A_{ij})
\label{eqH}
\end{equation}
where $\theta_i$ is the fluctuating phase of the superconducting wavefunction
on node $i$ of a periodic network of sites.  The sum is over pairs of
nearest neighbor sites, representing the bonds of the network, and
\begin{equation}
A_{ij}=(2\pi/\Phi_0)\int_i^j{\bf A}\cdot d{\bf\l}
\label{eqA}
\end{equation}
are fixed constants, giving
the integral of the magnetic vector potential across bond
$\langle ij\rangle$ ($\Phi_0=hc/2e$ is the magnetic flux quantum).
$U(\theta)$ is the interaction potential
between neighboring nodes, and its argument is just the gauge
invariant phase difference across the bond.  $U(\theta)$ is periodic
in $\theta$ with period $2\pi$, and has its minimum at $\theta=0$.
We will be interested here in the case of a uniform applied magnetic
field ${\bf\nabla}\times{\bf A}={\bf B}$, transverse to the plane
of the network.  In this case, the sum of the $A_{ij}$ going
counter clockwise around
any unit cell of the network is constant, and determined by
the magnetic flux through the cell,
\begin{equation}
\sum_{cell}A_{ij}=2\pi {\cal A}B/\Phi_0\equiv 2\pi f,
\end{equation}
where ${\cal A}$ is the area of a unit cell of the network.
$f$ therefore is the number of flux quanta of applied magnetic field,
per unit cell.

For an array of Josephson junctions, the interaction potential in
Eq.(\ref{eqH}) is taken as $U(\theta)=-J_0\cos(\theta)$.
For a superconducting wire network, in the London approximation, a
more appropriate interaction \cite{grest44}
is given by the Villain function,\cite{villain} defined by,
\begin{equation}
e^{-U(\theta)/T}\equiv\sum_{m=-\infty}^{\infty}
  e^{-J_0(\theta-2\pi m)^2/2T}
\label{eqV}
\end{equation}
where we take $k_B\equiv 1$.

For the Villain interaction, one can show by duality
transformation,\cite{jkkn}
that the Hamiltonian of Eq.(\ref{eqH}) can be mapped onto the following
2D classical Coulomb gas,
\begin{equation}
{\cal H} = {1\over 2} \sum_{ij} (n_i-f)V({\bf r}_i -{\bf r}_j)
				          (n_j-f),
\label{e3}
\end{equation}
where the sum is over all sites $i$, $j$ of the $dual$
lattice of the periodic network (i.e. the sites $i$ in Eq.(\ref{e3}) lie at the
centers of the unit cells of the network).  $n_i=0,\pm 1,\pm 2,...$
are integer ``charges" representing vortices in the phases $\theta_i$,
and the magnetic field flux density is represented by
the uniform background charge  $-f$.
$V({\bf r})$ is the lattice Coulomb potential in 2D, which solves
the equation,
\begin{equation}
\Delta^2 V({\bf r}) = -2\pi \delta_{\bf r,0},
\label{e4}
\end{equation}
where $\Delta^2$ is the discrete Laplacian for the network.
For large separations, $V({\bf r})\simeq \ln|{\bf r}|$.
In mapping from the network Hamiltonian given by Eqs.(\ref{eqH}) and
(\ref{eqV}), to the Coulomb gas Hamiltonian of Eq.(\ref{e3}),
we have followed convention\cite{minnrmp} by
rescaling the temperatures so that
$T_{CG}=T_{XY}/2\pi J_0$, where $T_{CG}$ refers to the temperature in the
Coulomb gas model, and $T_{XY}$ refers to the temperature in the
network (also referred to as a ``uniformly frustrated" $XY$ model\cite{tj}).
Henceforth, we will denote $T_{CG}$ as simply $T$.

Our simulations will be carried out in terms of this Coulomb gas problem,
rather than in terms of the phases $\theta_i$.
Although the Villain
interaction may give quantitative differences when compared to the
cosine interaction of a Josephson array, since the two functions have
the same symmetry, we expect that they will display the same qualitative
critical behavior.\cite{note}

For our simulations, we work with a finite $L\times L$ grid of sites,
and apply periodic boundary conditions to the Laplace Eq.(\ref{e4})
defining the Coulomb potential
$V({\bf r})$.  In this case, $V$ can be explicitly calculated in
terms of its Fourier transform.\cite{lee}  For a square network of lattice
constant $a_0$, one finds
\begin{equation}
V({\bf r})= {\pi\over N}\sum_{\bf k}
         {e^{i{\bf k}\cdot{\bf r}}\over 2-\cos({\bf k}\cdot{\bf a}_1)
				  -\cos({\bf k}\cdot{\bf a}_2)},
\label{e6}
\end{equation}
where $N=L^2$, $\{{\bf a}_1,{\bf a}_2\}$=
$\{a_0\hat{\bf x},a_0\hat{\bf y}\}$ are the basis vectors, and
the summation is over all wave vectors consistent with the periodic
boundary conditions, i.e. the set
$\{{\bf k}\} = \{(m_1/L){\bf b}_1 + (m_2/L){\bf b}_2\}$, with
$m_1,m_2 = 0,1,2 \dots L-1$, and with
$\{{\bf b}_1,{\bf b}_2\}=\{ (2\pi/a_0)\hat{\bf x},(2\pi/a_0)\hat{\bf y} \}$
the basis vectors of the reciprocal lattice.

For a honeycomb network, the charges $n_i$ sit on the dual triangular
grid of sites, and the Coulomb potential is given by,
\begin{equation}
V({\bf r})= {3\pi\over 2N}\sum_{\bf k} {e^{i{\bf k}\cdot{\bf r}}\over 3
        -\cos({\bf k}\cdot{\bf a}_1)
		-\cos({\bf k}\cdot{\bf a}_2)-\cos({\bf k}\cdot {\bf a}_3)},
\label{e5}
\end{equation}
where $\{{\bf a}_1,{\bf a}_2\}$=$\{a_0\hat{\bf x},a_0(\hat{\bf x}/2+
\sqrt 3\hat{\bf y}/2)\}$ are the basis vectors,
${\bf a}_3={\bf a}_2-{\bf a}_1$, and the wave vectors are determined by
$\{{\bf b}_1,{\bf b}_2\}=
\{(2\pi/a_0)(\hat{\bf x}-(1/\sqrt 3)\hat{\bf y}),
(2\pi/a_0)(2/\sqrt 3)\hat{\bf y}\}$.

The ${\bf k}=0$ terms in the summations of Eqs.(\ref{e6}) and (\ref{e5})
will cause a divergence in $V({\bf r})$. In real space, this is a
reflection of the
infinite self energy of a point charge.  Configurations with infinite total
energy will carry zero weight in the partition function sum, and may therefore
be excluded.  To keep the energy of the Coulomb gas
finite, we therefore impose the condition of overall charge neutrality
%7
\begin{equation}
\sum_i(n_i-f)=0.
\label{e7}
\end{equation}
If we define $N_c$ as the total number
of charges in the system, then Eq.({\ref{e7}) gives
\begin{equation}
N_c\equiv \sum_i{n_i}=fN.
\label{eqNc}
\end{equation}
Thus the density of magnetic flux quanta $f$, is equal to the density
of charges (vortices) $N_c/N$.
In the neutral system, the infinite self energies will exactly cancel, and
in place of $V({\bf r})$ we can
use only the nonsingular part of the Coulomb potential (\ref{e6}) and
(\ref{e5})
defined by $V'({\bf r})\equiv V({\bf r})-V({\bf r}=0)$. For a given system
size,
we evaluate $V'({\bf r})$ by numerically performing the summations indicated in
Eqs.(\ref{e6}) and (\ref{e5}).

The ground state will therefore be a
periodic vortex structure consisting of
$N_c$ sites with $n_i=+1$ (all other sites having $n_i=0$), spaced as
equally apart as allowed by the network geometry.  Understanding
the behavior of this
vortex structure at finite temperature will be one of the main goals of
this work.

%##############################################################################

\subsection{Relation to a uniform superconducting film}

The Coulomb gas model of the preceding section can also be used to describe
the melting of the vortex lattice in a uniform continuous superconducting
film.  For a superconducting film, the states of the system can be described
by a complex wavefunction, $\psi({\bf r})=|\psi({\bf r})|e^{i\theta({\bf r})}$.
As shown by Pearl,\cite{pearl}
for a of film of thickness $d$, provided the sample size is
smaller than the transverse magnetic penetration length $\lambda_\perp=
\lambda^2/d$, the magnetic field will  be essentially uniform and constant
throughout the film.  In this case, the states $\psi({\bf r})$ will be
weighted in the partition
function sum according to the Ginzburg-Landau free energy,
\begin{equation}
{\cal H}_{GL}[\psi]=\int d^2r \biggl\{\alpha |\psi|^2 +{1\over 2}\beta |\psi|^4
+
	 {1\over 2m}\biggl| \biggl({\hbar\over i}\nabla - {2e\over c}{\bf A}
                    \biggr) \psi\biggr|^2 \biggr\}
\label{eqF}
\end{equation}
with ${\bf\nabla}\times{\bf A}={\bf B}$ a fixed constant.
The mean field solution that minimizes ${\cal H}_{GL}[\psi]$, is similar to
that
found\cite{tinkham}
in three dimensions: ($i$) there is a triangular
lattice of equally spaced vortices in the phase $\theta({\bf r})$; ($ii$)
the areal density of vortices is
$B/\Phi_0$, with an average separation of $a_v\sim\sqrt{\Phi_0/B}$; ($iii$)
the size of the normal core of a vortex is determined by
$\xi_0\sim 1/\sqrt\alpha$, where $\alpha=0$ determines the $B=0$
mean field transition
temperature; ($iv$) the mean field phase transition at finite $B$ occurs
when $\xi_0\sim a_v$.

To include fluctuations, one should now sum the partition function over
all fluctuations of $\psi({\bf r})$ about the mean field solution.  In
doing so, one common approach has been to make the London approximation.
Here one
assumes that, outside of the normal vortex core, the amplitude of the
superconducting wavefunction is kept constant, and only the phase
fluctuates, i.e. $\psi({\bf r})=|\psi|e^{i\theta({\bf r})}$.
The London approximation is expected to be good whenever the bare
vortex core radius is very much smaller than the average
separation between vortices, $\xi_0\ll a_v$; by ($iv$) above,
this corresponds to
temperatures well below the mean field phase transition.

Substituting $\psi({\bf r})=|\psi |e^{i\theta({\bf r})}$ into Eq.(\ref{eqF})
results, within additive constants, in the simplified free energy
%2
\begin{equation}
{\cal H}[\theta]={1\over 2}J_0\int d^2 r \left|
    \nabla\theta -{2\pi\over\Phi_0} {\bf A}\right|^2,
\label{e2}
\end{equation}
where $J_0=\Phi_0^2/16\pi^3\lambda_\perp$,
and the integral is implicitly cut off at the vortex cores.
Eq.(\ref{e2}) is just a continuum version of the network Hamiltonian,
Eq.(\ref{eqH}).
Following Halperin and Nelson\cite{nhsuper} who considered the $B=0$ case, and
Huberman and Doniach\cite{hd} and Fisher\cite{fisher} who considered the
finite $B$ case, we note that Eq.(\ref{e2}) can be mapped onto a continuum
Coulomb gas of logarithmically interacting charges.  For finite $B$, this can
be written\cite{minnrmp} in the form of a one component
plasma on a uniform background charge density  $B/\Phi_0$,
\begin{equation}
{\cal H}={1\over 2}\int d^2 rd^2r^\prime [n({\bf r})-B/\Phi_0]V({\bf r}-
     {\bf r}^\prime)[n({\bf r}^\prime)-B/\Phi_0].
\label{eqHCGc}
\end{equation}
Here $n({\bf r})\equiv (1/2\pi){\bf \hat z}\cdot
{\bf\nabla}\times{\bf\nabla}\theta$ is the
vorticity in the phase of the superconducting wavefunction,
determined by singular integer vortices $n_i$ at positions ${\bf r}_i$,
$n({\bf r})=\sum_in_i\delta({\bf r}-{\bf r}_i)$.  $V({\bf r})$ solves the
2D Laplace equation, $\nabla^2 V=-2\pi\delta({\bf r})$.

The Coulomb gas of Eq.(\ref{e3}), introduced in the preceding section as a
description for a network, can now be viewed as a discrete approximation
to the continuum problem of Eq.(\ref{eqHCGc}).  For a fixed areal density
of vortices, $B/\Phi_0$, we recover the continuum Eq.(\ref{eqHCGc}) from
the discrete Eq.(\ref{e3}) as we take the network lattice constant $a_0\to 0$.
Since the number of vortices per unit cell in the network is
$f\sim a_0^2B/\Phi_0$, we see that the continuum is equivalent to the $f\to 0$
limit.  Thus by studying the melting of dilute vortex lattices in a network,
we can also learn about the melting of a vortex lattice in a uniform
superconducting film.
As in the previous section,
the mapping between the Coulomb gas and the superconductor is obtained by
measuring the Coulomb gas temperature $T_{CG}$ in units of $2\pi J_0=
\Phi_0^2 /8\pi^2\lambda_\perp$, i.e. $T_{super}=2\pi J_0T_{CG}$.

Finally, we note that the melting of the 2D vortex lattice, described by
the continuum Coulomb gas Hamiltonian of Eq.(\ref{eqHCGc}),
has been treated within the general 2D melting theory of
KTNHY.  Within
this theory, Fisher\cite{fisher}
has estimated that the melting transition occurs
more than an order of magnitude below the mean field transition.  This
observation completes the self consistency of the argument for using the
London approximation.

%#############################################################################

\subsection{Review of the theory of 2D melting}

	The analysis of our results will be guided by the ideas of the theory of
defect mediated melting in 2D, developed by KTNHY.\cite{ktmelt,nhmelt,young}
Although our results are, in many aspects,
in opposition to this KTNHY theory, it still represents a useful starting point
in exploring the phenomenon of 2D melting.

	For the 2D harmonic crystal on a smooth substrate (i.e. in the absence
of any  one-body potential) it is well known that fluctuations in the
long-wavelength phonon modes, lead to a logarithmic divergence in the
displacements of the particles, destroying translational long range
order at any finite temperature. This is a
consequence of the rigorous Mermin-Wagner theorem\cite{mermin} concerning
long range order in 2D. The standard theory
of elasticity shows however, that despite the absence of translational
long range order, the
low temperature phase of such a crystal is characterized by a slow power-law
decay of translational correlations,\cite{LLtext,nohex}
very different from the fast exponential
decay that one would expect in the liquid. This phenomenon has been termed
``quasi-long range'' order, and we shall refer to such a phase as a ``2D
solid''.  Based on the ideas of Kosterlitz and Thouless,\cite{ktmelt} that the
melting of such a 2D solid would be nucleated by the unbinding of topological
lattice defects, Nelson and Halperin\cite{nhmelt} and Young\cite{young}
formulated a theory (KTNHY) which predicted that 2D
melting would occur via two separate second order
KT-like transitions. In particular, they predicted that the 2D solid with
algebraic translational correlations  would
become unstable to the unbinding of dislocation defect pairs at
a temperature $T_m$, and melt into a new phase called the hexatic liquid.
This hexatic phase would be characterized by short range
translational order,
but quasi-long range six-fold orientational order. As the temperature
is increased, KTNHY predicted that this quasi-long range
orientational order would eventually be destroyed by the unbinding of
disclination defect  pairs, and at $T_i>T_m$, the hexatic liquid would
melt into a normal (isotropic) liquid with short range orientational order.
We summarize this scenario by
writing down the long-range limiting behavior predicted for
the translational and orientational correlation functions.

For translational correlations,
%12
\begin{equation}
	\langle e^{i\bG\cdot(\br_i-\br_j)}\rangle
             \sim\left\{
              \begin{array}{ll}
               1   		     & {\rm in\ perfect\ crystal}\ (T=0) \\
	       r_{ij}^{-\eta_\bG(T)} & {\rm in\ 2D\ solid}\ (0<T<T_m)\\
         e^{-r_{ij}/\xi_+} \;  & {\rm in\ hexatic\ or\ normal\ liquid}\ (T>T_m)
              \end{array}
         \right.
\label{e11}
\end{equation}
where $r_{ij}=|\br_i-\br_j|$ is the separation between particles $i$ and $j$,
and $\bG$ is a reciprocal lattice vector of the
perfect (triangular) crystal at $T=0$.
$\eta_\bG(T)$ is a temperature dependent
exponent, which for the 2D harmonic crystal can
be expressed in terms its shear modulus $\mu$ and bulk modulus $\lambda$, as
\begin{equation}
\eta_\bG(T)={k_B T |\bG|^2(3\mu +\lambda)\over 4\pi\mu(2\mu +\lambda)}.
\end{equation}
In the 2D vortex lattice, the bulk modulus $\lambda$ is infinite because of
the long-ranged nature of the Coulomb interaction. The expression for
$\eta_\bG(T)$ thus simplifies to,
\begin{equation}
\eta_\bG(T)={k_B T |\bG|^2\over 4\pi\mu}.
\label{e12}
\end{equation}
A key prediction of the
KTNHY theory is that if $\bG_1$ is a shortest reciprocal lattice vector, then
$\eta^{-1}_{\bG_1}(T)$ takes a discontinuous jump  at $T_m$ to zero from the
universal value,
\begin{equation}
\eta^{-1}_{\bG_1}(T^-_m)=3.
\label{eqetau}
\end{equation}
In what follows, we will directly test
this prediction. We wish to stress however, that the
behavior of translational
correlations in the 2D solid phase, as given by Eq.(\ref{e11}), is a general
result of continuum elastic theory,
independent of all assumptions concerning the mechanism of the melting
transition. It is only the universal
jump in $\eta^{-1}_{\bG_1}(T_m)$, and the existence of the hexatic phase,
which are specific predictions of KTNHY.

	The six-fold orientational correlation function, according to KTNHY,
behaves as
%13
\begin{equation}
	\langle e^{6i[\vartheta(\br_i)-\vartheta(\br_j)]}\rangle
             \sim\left\{
              \begin{array}{ll}
        \alpha e^{-r_{ij}/\xi'_6}+\phif\;&{\rm in\ 2D\ solid}\ (0<T<T_m)\\
r_{ij}^{-\eta_6(T)}             & {\rm in\ hexatic\ liquid}\ (T_m<T<T_i)\\
        e^{-r_{ij}/\xi_6}               & {\rm in\ normal\ liquid}\ (T>T_i)
              \end{array}
              \right.
\label{e13}
\end{equation}
where $\vartheta(\br_i)$
is the angle of the bond from particle $i$ to its neatest neighbor, relative to
some fixed reference direction.
$\alpha$ is a proportionality constant of order one, and $\phif$ gives the
value of the long range orientational order expected in the 2D solid phase.
The exponent $\eta^{-1}_6(T)$, describing the quasi-long range order of
the hexatic phase, is predicted to have a universal jump
to zero at $T_i$ from the value $\eta^{-1}_6(T_i^-)=4$.

In the 2D solid, a relation
between $\phif$ and the vortex shear modulus $\mu$  can be derived from
continuum elastic theory,\cite{nohex}
\begin{equation}
\phif \simeq \exp\biggl[-{9k_B T \Lambda^2\over 8\pi\mu} \biggr]
=\exp\biggl[-{9\Lambda^2\eta_\bG\over 2|\bG|^2}\biggr],
\label{e14}
\end{equation}
where $\Lambda\sim 2\pi/a_v$ is an ultraviolet cutoff ($a_v$ is the average
separation between particles). Since we will
independently measure $\phif$ and $\eta_\bG$ in our simulation, we will use
this relation as a check of the consistency of our results.

For a periodic superconducting network, we have discussed how the discrete
substrate of the network serves to induce a periodic pinning potential for
the magnetic field induced vortices.
To treat this case, we are therefore interested in how the
above 2D melting scenario is altered by the presence of a periodic substrate.
We shall be interested in the
situation where the period of the substrate is sufficiently small compared to
the spacing between particles, so that the essential
features of the defect mediated melting theory remain intact.
This problem has been treated by Nelson and Halperin\cite{nhmelt}.
The main result of such a
``fine-mesh" periodic perturbation is the appearance of a new phase at
low temperatures,
in which the 2D solid is commensurably pinned to the substrate.
This phase has true long range translational order,
and we shall refer to it as  the ``pinned solid''. At
a certain depinning temperature $T_c<T_m$, there is a transition
to a 2D  ``floating solid'' phase, where the solid decouples from the
substrate,
and translational correlations behave identically to
those of a 2D solid on a uniform substrate; this triangular floating solid
may in general be incommensurate with the periodic substrate.
Increasing temperature, the
floating solid is expected to melt at $T_m$,
via the dislocation unbinding mechanism, into a
liquid phase. On a triangular substrate, this liquid will have a small
(but finite) long-ranged six-fold orientational order induced by the substrate,
at all temperatures. There should, however, be a temperature $T_i$ where
$\phif(T)$ shows a significant drop,
reminiscent of the disclination unbinding transition
on the smooth substrate. This drop should become increasingly sharper
as the ratio of substrate period to particle separation becomes smaller.
On the square substrate, Nelson and Halperin predict that there will be
a sharp Ising transition at a $T_i>T_m$, where quasi-long range
six-fold orientational order in the liquid vanishes,
and only the long range four-fold orientational order induced
by the substrate remains.
This Ising transition can be viewed as a ``ghost'' of the hexatic to normal
liquid transition, which would occur in the absence of the periodic substrate.
The four-fold orientational order, induced by the substrate, again persists
at all higher temperatures.

To conclude, we note again that the properties of the 2D floating solid,
described above, follow solely from continuum elastic theory, independent
any particular theory of melting.  It is the existence of the hexatic
liquid phase that is a specific prediction of the KTNHY melting theory.
However,  as pointed out by Nelson and Halperin,\cite{nhmelt} it is always
possible that a
``premature" unbinding of disclination pairs may lead to a direct melting of
the
2D solid into the normal liquid. Such a transition is then expected to be first
order.  In this case, the KTNHY prediction, Eq.(\ref{eqetau}), for the
universal
value of $\eta^{-1}_\bG$ at
melting, becomes a lower bound, $\eta^{-1}_{\bG_1}(T_m^-)\ge 3$.
Results from various numerical simulations and experiments\cite{nohex}
indicate that this first order behavior might indeed be prevailing in the
various 2D systems studied so far.

%############################################################################

\subsection{Observables and finite size scaling}

	We now show how the predictions of the preceding section
translate into the behavior
of observables  which can be directly measured in our MC simulation.
There are two key issues that we wish to investigate in the
superconducting networks: ($i$) the transition from the superconducting
to the normal state, and ($ii$) the
melting of the magnetic field induced vortex lattice.  For ($ii$), our
goal is to test the KTNHY theory of 2D melting, and so we will be interested
in studying both the translational and the orientational order of the vortex
lattice.

The superconducting to normal transition, marked by the loss of superconducting
phase coherence, is measured
by the vanishing of the {\em  helicity modulus}, $\Upsilon(T)$,
which measures the response
of the system to applying a net twist, or phase gradient, to the
phases $\theta_i$ in the Hamiltonian, Eq.(\ref{eqH}).  For the Villain
interaction of Eq.(\ref{eqV}), the helicity modulus can be shown\cite{ohta}
to be identical to the
inverse dielectric function of the corresponding Coulomb gas of Eq.(\ref{e3}),
$\Upsilon/J_0=\epsilon^{-1}$, where $\epsilon^{-1}$ is defined
in the usual way,
\begin{equation}
\epsilon^{-1}(T)=\lim_{k\rightarrow 0}\biggl\{1-{2\pi\over TNk^2}
                 \langle n_{\bf k} n_{-\bf k} \rangle \biggr\}.
\label{e8}
\end{equation}
Here $n_{\bf k}=\sum_i n_i\exp(-i {\bf k}\cdot {\bf r}_i)$ is the
Fourier-transformed charge density.  The vanishing of $\epsilon^{-1}$
signals an insulator to metal transition in the Coulomb gas.  The
free charges characteristic of the conducting phase correspond to
freely diffusing vortices in the superconducting network, which are
responsible for the loss of phase coherence.\cite{ktmelt}
In the simulation, the ${\bf k}\rightarrow 0$ limit is approximated
by averaging $\epsilon^{-1}$ over the smallest allowed nonzero wave vectors.

	Information on the translational order in the vortex lattice can be
extracted from the {\em structure function}
%9
\begin{equation}
S({\bf k})={1\over N_c}\langle n_{\bf k} n_{-\bf k} \rangle
           \equiv{1\over N_c}\sum_{ij}e^{i{\bf k}\cdot ({\bf r}_i-{\bf r}_j)}
           \langle n_{i} n_{j} \rangle,
\label{e9}
\end{equation}
which we evaluate for all allowed wave vectors
$\bk=(m_1/L){\bf b}_1+(m_2/L){\bf b}_2$ in the first Brillouin
zone (BZ) of the reciprocal lattice to the real space dual lattice of the
superconducting network.
A 2D intensity plot of $S(\bk)$ serves as a simple tool for
visualization of the different phases in the system. In analogy to the
conventional X-ray scattering images, we expect $S(\bk)$ to display a periodic
array of sharp delta-function Bragg peaks in a state with long range
translational
order, and a set of smooth concentric rings in a normal liquid phase. A
phase with quasi-long range translational order, characterized by
algebraic translational correlations, will be distinguished
by a regular array of algebraically diverging peaks of finite
width.\cite{nohex}
A hexatic liquid phase should appear as a set of concentric
rings with six-fold angular modulation.

Apart from providing the simple visualization described above, the
scaling of the heights of the peaks in $S(\bk)$,
 as a function of system size $L$,
will serve as a good quantitative indicator
of translational correlations in the system.
Combining the definition of $S(\bk)$ in Eq.(\ref{e9}) with Eq.(\ref{e11})
(note that in Eq.(\ref{e9}), $n_i=1$ on a site containing a vortex and
$n_i=0$ on a site without a vortex) one easily obtains
%\addtocounter{equation}{1}
%\newlabel{e15}{{\value{equation}}{6}}
\begin{equation}
		{S(\bG)\over L^2}
\sim\left\{
\begin{array}{llr}
1                 & {\rm in\ pinned\ solid}\ (T<T_c)              &
\;(\arabic{equation}{\rm a})\\
L^{-\eta_\bG(T)}\;& {\rm in\ floating\ solid}\ (T_c<T<T_m)        &
\;(\arabic{equation}{\rm b})\\
(\xi_+/L)^2       & {\rm in\ hexatic\ or\ normal\ liquid}\ (T>T_m)&
\;(\arabic{equation}{\rm c})
\end{array}
\right.
\label{e15}
\end{equation}
The finite size scaling analysis of the translational order, that we present in
Section III, will be based on the above relations. In particular, a comparison
of Eqs.(\ref{e15}) with our MC data will allow us to extract the temperature
dependent exponent $\eta_\bG(T)$ and test the KTNHY prediction regarding
the universal jump in $\eta^{-1}_{\bG_1}(T_m^-)$.
We shall also determine the correlation length $\xi_+(T)$ in the liquid phase.

	There is an independent way to extract the exponents $\eta_{\bG}$
(and thus the
vortex shear modulus $\mu$) without having to use finite size scaling.
One can instead, for a given system size, fit to the heights of the peaks
$S(\bG)$, as a function of $|\bG|$.
For the low order peaks, this dependence is roughly Gaussian, as can be seen
by combining Eq.(\ref{e15}b) with the expression for $\eta_\bG$ in
Eq.(\ref{e12}).
For the higher order peaks however, we need to rederive this dependence,
since the
prefactor (which is not shown in Eq.(\ref{e15}b)) becomes important.
Substituting
Eq.(\ref{e11}) into the definition of the structure function, Eq.(\ref{e9}),
and approximating the summations by integrations, we get
%21.1
\begin{equation}
S(\bG)=c\int d^2r (r/a_v)^{-\eta_\bG} \simeq
	      2\pi c a_v^{-\eta_\bG} \int_{a_v}^R  dr r^{1-\eta_\bG},
\label{e21.1}
\end{equation}
where $R\sim L$ is a long distance cutoff, $a_v$ is the average separation
between vortices, and $c$ is a proportionality constant
of the order one. The integral is easily evaluated
\begin{equation}
S(\bG)/L^2 \simeq {2\pi c\over
2-\eta_\bG}\biggl[(R/a_v)^{-\eta_\bG}-(R/a_v)^{-2}
     \biggr],
\label{e21.2}
\end{equation}
and $c$ is determined by the requirement that $S(0)/L^2=1$. One can see that
for $\eta_\bG \ll 2$ formula (\ref{e15}b) remains valid, but for $\eta_\bG
\simeq 2$ it breaks down. In practice, since $\eta_{\bG_1}<1/3$ by the
KTNHY bound, and any
exponent $\eta_\bG$ at a given fixed temperature can be written as
\begin{equation}
\eta_\bG= \eta_{\bG_1}\cdot(|\bG|^2/ |\bG_1|^2),
\label{e21.3}
\end{equation}
formula (\ref{e15}b) will approximately hold for the three shortest reciprocal
lattice
vectors $\bG$ only. To draw quantitative conclusions regarding the exponent
$\eta_\bG$,
one must use the more accurate relation of Eq.(\ref{e21.2}).

Information on the bond orientational order will be obtained
by measuring the four-fold and six-fold {\em orientational correlation }
%10
\begin{equation}
\varphi_p(T)={1\over N^2_c}\sum_{ij}\biggl\langle
              e^{ip(\vartheta_i-\vartheta_j)}\biggr\rangle, \ p=4,6
\label{e10}
\end{equation}
where the sum is over sites with non-vanishing charges $n_i=+1$, and
$\vartheta_i$
is the bond orientation angle defined in the previous section.
For a finite system, one expects to see a sharp drop
in $\varphi_p(T)$ at the transition from an orientationaly ordered phase,
to a disordered or possibly hexatic phase.
One can deduce the scaling of the orientational correlation
$\varphi_6(T)$ with system size $L$, by combining Eq.(\ref{e10}) with the
KTNHY prediction of Eq.(\ref{e13})
%
%{\addtocounter{equation}{1}
%\newlabel{e16}{{\value{equation}}{6}}
\begin{equation}
\varphi_6\sim\left\{
\begin{array}{llr}
2\pi\alpha (\xi'_6/L)^2+\phif &{\rm in\ pinned\ or\ floating\ solid }\
(0<T<T_m)
 &\;(\arabic{equation} {\rm a})\\
L^{-\eta_6(T)}           & {\rm in\ hexatic\ liquid}\ (T_m<T<T_i)        &
\;(\arabic{equation}{\rm b})\\
(\xi_6/L)^2              & {\rm in\ normal\ liquid}\ (T>T_i)             &
\;(\arabic{equation}{\rm c})
\end{array}
\right.
\label{e16}
\end{equation}
Relations (\ref{e16}a) and (\ref{e16}c) hold for $L\gg\xi_6$  (otherwise
one must include corrections
$\sim \exp(-L/\xi_6)$). These scaling relations for  $\varphi_6$ will be
used extensively in our analysis, to test for the existence of a hexatic phase
in our model.

%%%%%%%%%%%%%%%%%%%%%%%%%%%%%%%%%%%%%%%%%%%%%%%%%%%%%%%%%%%%%%%%%

\subsection{Monte Carlo algorithm}

	For the purpose of developing a fast MC algorithm, it is important to
realize that
the physical phenomena described in the previous section occur at temperatures
which are about one order of magnitude lower\cite{fisher} than the ordinary
Kosterlitz-Thouless (KT) transition\cite{nhsuper}
in the zero magnetic field, $f=0$, case.
This implies that the role of vortex-antivortex
pair excitations is negligible in the temperature range that we study,
and that in the simulation we can restrict ourselves
to the excitations caused by movement of the vortices induced by the external
magnetic field ${\bf B}$. We have
explicitly verified that the energy of an isolated vortex-antivortex pair (in
the
Coulomb gas language a pair of $(+,-)$ charges) is always $E_{pair}\gg k_BT$,
and thus in
practice such an excitation would never be accepted in the MC simulation.
Consequently,
our updating scheme is as follows. In each step, one charge is selected at
random
and moved to a different site within a radius $r_0$, which is chosen so as
to maximize the acceptance rate. We find that values $r_0\sim a_v/2$ are
optimal.
The energy change $\Delta E=E_{new}-E_{old}$ is then computed, and the
excitation
is accepted or rejected according to the standard Metropolis algorithm:
\[
	{\rm accept\ if\ }\; e^{-\Delta E/T} > x,
\]
where $x$ is a random number uniformly distributed on the interval $[0,1)$.
Here and henceforth, we work in units in which $k_B\equiv 1$.
$N_c$ such attempts we will refer to as one MC
sweep. At low temperature, we also made global moves, by attempting to shift
entire rows of charges by one space.  Such moves are meant to model long
wavelength shear excitations, and help to accelerate equilibration near
the vortex lattice melting transition.

	Due to the long-ranged nature of the Coulomb potential, the most time
consuming
operation is the evaluation of $\Delta E$. From Eq.(\ref{e3}) we find that the
energy
change for moving a charge from the site $\bR_1$ to the site $\bR_2$ is
\begin{equation}
\Delta E=-\sum_j V'(\bR_1-\br_j) n_j +\sum_jV'(\bR_2-\br_j) n_j -
V'(\bR_1-\bR_2),
\label{e18}
\end{equation}
where we have used the fact that $V'(-\br)=V'(\br)$ and $V'(0)=0$. In this
form,
each evaluation of $\Delta E$ is a computation of the order $N_c$, as $j$
sweeps
through all the sites with nonvanishing charge $n_j=+1$. To speed up this
process,
we use an algorithm developed by Grest.\cite{grest}  At each site of the
lattice
we define a potential due to all charges in the system
\begin{equation}
F(\br_i) = \sum_j V'(\br_i-\br_j) n_j.
\label{e19}
\end{equation}
Now each evaluation of
\begin{equation}
\Delta E = -F(\bR_1) + F(\bR_2) - V'(\bR_1-\bR_2)
\label{e20}
\end{equation}
requires computation of only $O(1)$. Naturally, each time the excitation is
accepted, it is necessary to update $F(\br_i)$ at all sites
\begin{equation}
F_{new}(\br_i) = F_{old}(\br_i) - V'(\br_i-\bR_1) + V'(\br_i-\bR_2),\;
i=1,2,\dots N.
\label{e21}
\end{equation}
This is a computation of order $N$. However, since the acceptance rate in the
interesting temperature range is very low (typically below 1\%), this method
is faster than the direct approach of Eq.(\ref{e18}).

Data is collected by
heating the system up from the ground state. At each temperature
we discard $30,000$ MC sweeps to equilibrate the system. Then, starting from
this equilibrated configuration, we perform several (typically $4-6$)
independent
runs of $100,000$ sweeps each to sample physical quantities. In some cases,
when evaluating quantities at the temperatures close to the critical
point, substantially longer runs are carried out.
 Errors are estimated
from the standard deviation of these independent runs.
To verify the consistency of our results, we also perform cooling from a random
configuration at high temperature; no substantial hysteresis is found.

	All simulations were carried out on Sparc 10 workstations. The time
needed to equilibrate the system and sample the physical quantities at a
given temperature $T$ was typically several hours (depending on size) using
100\% of the single processor power. For example, it took approximately
3 hours to carry out 100,000 MC sweeps at a temperature close to melting,
for a medium-sized system $N_c=81$ with density $f=1/49$. Our longest run, to
sample the energy distribution near the depinning transition, took 189 hours
for $4\times 10^7$ MC sweeps on  the largest system of  $N_c=169$ and $f=1/49$.

\section{ Simulations on the triangular grid: dilute case}

\subsection{The results}

	In this section we report our results from simulations of the
Coulomb gas Hamiltonian
(\ref{e3}) on a triangular grid of sites (corresponding to a
honeycomb superconducting network), for the dilute limit
$f\ll 1$ (or equivalently $a_0\ll a_v$).
In this case, we expect that
our discretized model
will well approximate the continuum.  Some of these results have been
reported by us previously.\cite{FrTe}
The advantage of choosing a triangular
grid is that, for a given system size $L$, one can always choose $f$ in such
a way so as to accommodate a perfect, commensurate,
triangular vortex lattice in the ground state.
By contrast, this is never possible on a square grid.
It is convenient to
choose $f=1/m^2$, with $m$ integer, since then each system size of the form
$L=s\cdot m$, ($s$ integer), will accommodate a triangular ground state with
$N_c\equiv f L^2=s^2$ vortices. We have studied systematically densities
$f=1/m^2$, with $m=3$ to $12$, and fixed $N_c\simeq 100$. The results of our
investigation are summarized in Figure \ref{f1}: for sufficiently dilute
systems ($f<1/25$) we find three distinct phases. At low temperatures the
vortex
lattice is in a ``pinned solid" phase, locked to the underlying grid.
Above a sharp depinning temperature $T_c(f)$, the vortices are in a
``floating solid" phase, which then melts at $T_m$ into a normal vortex liquid.
The properties of these phases will be discussed below. For denser systems with
$f>1/25$, the two transitions at $T_c$ and $T_m$
merge, and there is only a single transition from a
pinned solid into a liquid.

	For a simple visualization of the three phases, we show in Figure
\ref{f2}
intensity plots of $S(\bk)$ at various $T$,
for the specific case of $f=1/49$ and $N_c=63$.
We also display the amplitude of $S(\bk)$ along the symmetry axis $k_y$.
For $T=0.003$ (Figure \ref{f2}a), just below $T_c(f)$, we see a regular array
of $\delta$-function Bragg peaks, indicating long ranged translational order
induced by pinning to the triangular grid. The width of these peaks corresponds
to the
finite resolution of wave vectors allowed by our finite system. At $T=0.0065$
(Figure \ref{f2}b), just bellow $T_m$, we see a regular array of peaks, but
they
are now of finite width. We will show that these peaks are consistent with
the power law
singularities characteristic of the algebraic translational correlations
expected for a 2D floating solid phase. The heights of the peaks along the
symmetry axis are well described by a Gaussian, as expected from
Eqs. (\ref{e21.2})
and (\ref{e21.3}). Thus, for $T_c(f)<T<T_m$ we do have a floating vortex
lattice,
as in the continuum limit. For $T=0.0075$ (Figure \ref{f2}c),
slightly above $T_m$ we see a rotationaly invariant structure, typical for a
liquid with short range correlations. Thus for $T>T_m$,
the floating vortex lattice has
melted into a normal liquid. It is interesting to note that we see
no sign of angular modulation in the rings above $T_m$. One might expect such a
modulation due to the long ranged six-fold orientational order
induced in principle by the
underlying triangular grid; if the grid was too fine for this effect to be
significant, modulation might still be present if a
hexatic phase existed just above $T_m$.

	In Figure \ref{f3a}, we plot versus $T$ the inverse dielectric function
$\epsilon^{-1}(T)$, and the orientation order correlation  $\varphi_6(T)$,
for $f=1/49$ and $N_c=169$ (one of the largest systems that we have studied).
We see that $\epsilon^{-1}(T)$ vanishes at the depinning transition $T_c(f)$,
signaling the loss of superconducting phase coherence in the floating solid
phase.
This is just a reflection of the fact that an
unpinned vortex lattice, in the presence of any applied d.c. current (no matter
how small), will be free to drift transversely to the
current, resulting in a finite linear ``flux flow" resistance.
Our results explicitly
show that the absence of phase coherence in this $\bk\rightarrow 0$ sense, does
not imply the absence of a well defined vortex lattice.
Considering the orientation order, we see that $\varphi_6$ sharply
drops at $T_c(f)$, but remains finite up to the melting temperature $T_m$,
where it drops again sharply to nearly zero values. The smallness of
$\varphi_6$
above $T_m$ indicates that the six-fold orientational long range order
which is induced in principle by the
triangular grid, is indeed a negligibly small effect at the densities we are
concerned with. We shall discuss this point in more detail in the following
section.
In the  Figure \ref{f3b}, we show the dependence of $T_c(f)$ and $T_m$ on the
vortex density $f$, as estimated from the behavior of $\epsilon^{-1}(T)$ and
$\varphi_6(T)$, and checked against the behavior of the structure function
$S(\bk)$.
We see that only for sufficiently dilute systems, $f<1/25$,
is there a floating solid phase; for $f>1/25$ there is only a single
transition from a pinned solid to a liquid.  As $f$ decreases, $T_c(f)$
vanishes linearly with $f$, consistent with the TJ conjecture\cite{tj} for the
loss
of superconducting coherence.
$T_m$ however, quickly approaches a finite constant
$T_m=0.0070\pm0.0005$.
In terms of the superconductor temperature, this means a vortex
lattice melting at $T_m=0.0070\>\Phi_0^2/8\pi^2\lambda_\perp$.  This is
well within the bounds $0.0046<T_m<0.0086$ estimated by Fisher\cite{fisher}
from the KTNHY theory.


\subsection{The melting transition: finite size scaling analysis}

	To investigate if the melting transition at $T_m$ is indeed consistent
with the KTNHY
theory, we have carried out a detailed finite size scaling analysis for the
density $f=1/49$. This density has been chosen for two reasons. First,
the estimated $T_m$ is well separated from the depinning temperature $T_c$,
and hence the floating solid phase exists in a relatively wide interval
of temperatures. Second, the density is not too small, and thus we are
able to study systems with as many as $N_c=169$ vortices.
More dilute systems, with comparable
$N_c$ would require sizes that are currently out of reach of the computer power
available to us. We have carried out extensive simulations for the system sizes
$L=28,35,\dots 91$, and we have analyzed the size-dependencies of various
physical
quantities at several temperatures below and above $T_m$. Our results are as
follows.

In Figure \ref{f4a} we plot $S({\bf G}_1)/L^2$
as a function of $L$, on a log-log scale, for several
different temperatures.  Data for each temperature fall on
a straight line, confirming the expected power-law behavior of Eq.(\ref{e15}).
These straight lines fall into three distinct groups.
For $T<T_c\simeq 0.0045$,
$S({\bf G}_1)/L^2\sim 1$, indicating the long range order of the pinned
lattice.
For $T_c<T<T_m\simeq 0.007$, we find algebraic decay,
$S({\bf G}_1)/L^2\sim L^{-\eta_{\bf G}(T)}$.  For $T>T_m$, we find
$S({\bf G}_1)/L^2\sim L^{-x}$, with $x\to 2$ as $T$ increases,
consistent with the short range order of a liquid.

Thus, our data for the floating solid phase are consistent with the
predictions of the continuum elastic theory, given by Eq.(\ref{e15}b),
and in particular we may fit our data to this expression to obtain the
translational correlation exponent $\eta_{\bG_1}$.  We show our results
in Table \ref{t1}.  We can now make quantitative comparison with the KTNHY
theory, by noting that $\eta_{\bG_1}$ first exceeds the
KTNHY universal value of $1/3$ (see Eq.(\ref{eqetau})
at $T=0.0065$, very close to the melting
temperature $T_m\simeq 0.0070$ as estimated from the behavior of the
orientational
correlation $\varphi_6(T)$ of Figure \ref{f3a}.
The slopes of the lines in Figure \ref{f4a} also show an apparent
discontinuous jump at this same $T_m$.

As a consistency check, we have also computed $\eta_{\bf G_2}$, where
${\bf G}_2=2{\bf G}_1$.
Using similar fits to $S(\bG_2)$ as in Figure \ref{f4a}, we determine the
exponent
$\eta_{{\bf G}_2}$, and show the results in Table \ref{t1}.
We see that $\eta_{{\bf G}_2}\simeq 4\eta_{{\bf G}_1}$ as expected,
since according to Eq.(\ref{e14}) $\eta_{\bf G}\sim |{\bf G}|^2$.

As an alternative way of calculating $\eta_{\bG_1}$,
we fit to the heights of the peaks in $S(\bk)$ at all the available $\bG$,
for a fixed size system, as described in Eqs.(\ref{e21.1}) $-$ (\ref{e21.3}).
We found that the results
are only weakly dependent on the precise value of the cutoff $R$
of Eq.(\ref{e21.2}); we therefore take
$R=L$, which results in an excellent fit. We show one example of such a fit
in Figure \ref{f6}. The exponent $\eta_{\bG_1}(T)$, obtained in this way, is
shown
in Table \ref{t2} for the sizes $L=63,77,91$, with $f=1/49$.
We note, that
despite a certain tendency to overestimation, these
exponents are in  reasonable agreement with those obtained from the finite
size scaling.
This method of extracting the shear modulus should be useful in
situations where a finite size scaling analysis is difficult to handle, such as
in systems with a large unit cell in the ground state.

	Let us now consider the orientational order. In Figure {\ref{f4b}} we
plot the orientational correlation $\varphi_6(T)$ as a function of $L$ for
several
temperatures. In the pinned solid, $T<T_c$, $\varphi_6(T)\rightarrow 1$ as
$L$ increases, confirming the expected long-ranged orientational order of the
perfect pinned triangular lattice. More interestingly, $\varphi_6(T)$ also
approaches
a finite value $\phif$ in the floating lattice phase, $T_c<T<T_m$, in agreement
with
continuum elastic theory. The solid lines in Figure {\ref{f4b}}
are from least squares fits to Eq.(\ref{e16}a).
The resulting fitted values of $\phif$
are shown in Table \ref{t1}. Above $T_m$ we attempt to fit to the power law of
Eq.(\ref{e16}b) for a hexatic liquid, but we always find that using
Eq.(\ref{e16}c) for an isotropic liquid, results in a distinctly better fit.
Since the underlying triangular grid will in principle result in long range
six-fold order at all temperatures, we have also fit our data above $T_m$ to
the form of Eq.(\ref{e16}a), which differs from Eq.(\ref{e16}c) only in the
constant $\phif$.  As shown in Table \ref{t1} however, we always find
$\phif\simeq 0$.  Thus the discrete grid is playing a negligible role in the
orientation order.  To compare our fits above $T_m$, we note that
in most cases the $\chi^2$-parameter
of the fit to Eq.(\ref{e16}a) is 5 to 10 times smaller than that of
Eq.(\ref{e16}b).
The former fit is also much more stable in the sense that fitted parameters
do not
change significantly when the data is restricted to different ranges of $L$.
We may therefore conclude that, in agreement with
our investigation of the structure function, the floating solid melts directly
into a normal liquid.  The hexatic phase is either absent in our system, or it
occurs only in some extremely narrow interval of temperatures, which
makes it difficult to detect by numerical simulation.

In order to see this another way, in Figure \ref{f5a}
we plot versus temperature our values of $\phif(T)$ and $T/\eta_{\bG_1}(T)$,
obtained from our finite size scaling analysis.  From Eq.(\ref{e12}) we see
that $T/\eta_{\bG_1}(T)$ is just proportional to the vortex lattice shear
modulus $\mu$.  We see, that $\phif(T)$ starts to drop
at the same temperature that $T/\eta_{\bG_1}(T)$ first drops below
the KTNHY universal value of $3T$ (see Eq.(\ref{eqetau})),
i.e. the temperature at which  the floating solid starts to melt.
The temperature range over which $\phif$ decays to zero is identical  to
the range over which  $T/\eta_{\bG_1}(T)$ decays. This suggests
that the small but finite values of $\phif$  which we find
above $T_m$ are just a finite size effect, rather than a signature of the
hexatic phase.
Let us also note that the exponent  $\eta_{\bG_1}(T)$ has a physical meaning
only
below $T_m$. Above $T_m$ it is strictly infinite and we use it here, with some
abuse of notation, simply as the exponent resulting from the fit of our data to
the power law form of Eq.(\ref{e15}b).

	Having obtained the values of  $\eta_{\bG_1}(T)$ and $\phif(T)$,
we can now test the relation, Eq.(\ref{e14}), between the orientational
long range order and the vortex lattice shear modulus $\mu$, that should hold
in the floating solid phase. Expressing $\mu$ in terms of the exponent
$\eta_{\bG_1}$, Eq.(\ref{e14}) gives
%23
\begin{equation}
\eta_{\bG}(T)=-K \ln[\phif(T)],
\label{e23}
\end{equation}
with $K=2|\bG|^2/9 \Lambda^2$, where $\Lambda =\lambda(2\pi/a_v)$,
and $\lambda$ is a dimensionless constant of order unity.
In Figure \ref{f5b} we plot
$\eta^{-1}_{\bG_1}(T)$ and $-1/K \ln[\phif(T)]$ versus temperature. For
$K=0.33$,
which corresponds to $\lambda=0.94$, the two data sets lie on top of each other
for all $T$ below $T_m$, providing yet another consistency check for our
calculation.

	By fitting our data above $T_m$ to Eqs.(\ref{e15}c) and (\ref{e16}c)
we have
also extracted the correlation lengths $\xi_+(T)$, associated with
translational
order, and $\xi_6(T)$, associated with six-fold bond orientational order.
We are able to determine these from finite size scaling only up to an
overall multiplicative factor. We determine this factor by assuming  that at
high $T$, the correlation lengths are equal to the average spacing between
vortices,
i.e. $\xi(T\rightarrow \infty) = a_v$. With this assumption, $\xi_+(T)$ and
$\xi_6(T)$ are displayed in Figure \ref{f7}. We see that both correlation
lengths
rapidly increase around $T\simeq 0.007$. It is also evident from Figure
\ref{f7},
that orientational correlations persist out to longer distances,
up to higher temperatures, than translational correlations.


\subsection{ The order of melting transition}

 The absence of the hexatic phase, as deduced from our analysis of the
orientational correlations, suggests the possibility that
the transition is not of the KTNHY type, but is due to some other mechanism,
such as domain wall proliferation. It might also be, that the
unbinding of disclinations occurs simultaneously with the unbinding of
dislocations.
Such a possibility has been suggested in Ref.\onlinecite{nhmelt}. In any case,
it is useful to determine  the order of this melting transition.
To examine the possibility that the transition is first order,
we have used the histogram method due to Lee and
Kosterlitz \cite{hist}. For various system sizes at $f=1/49$, we measure the
energy
distribution $P(E)\sim e^{-F(E)/T}$ near the melting temperature $T_m$.
In Figure \ref{f8} we plot the resulting free energy $F(E)$ versus $E$.
Although
our data are somewhat noisy, we see a clear double well structure with an
energy barrier $\Delta F$ between two coexisting phases.  The inset to
Figure \ref{f8} shows the dependence of $\Delta F$ on the system size $L$.
The energy barrier $\Delta F$ grows with $L$, strongly suggesting a first
order transition. Our system sizes remain too small to see clearly the
predicted
scaling $\Delta F \sim L$.

	For all sizes, the data have been taken at $T=0.0065$, and then the
energy distribution is extrapolated, using the method of Ferrenberg
and Swendsen,\cite{extrapolate} to that temperature which gives two minima
of equal depth.
This criterion gives an improved estimate of the melting temperature,
$T_m=0.0066$.
A total of $10^7$ MC sweeps were performed for each size to measure the energy
distribution $P(E)$. We have checked the consistency of these measurements by
calculating the energy of the system at various temperatures (above and below
$T_m$) using the extrapolated distributions $P(E,T)$.
We then compared these with
energies obtained by direct simulation at those temperatures, and found  good
agreement for all temperatures not too far from $T_m$.

	To conclude, the histogram method provides strong evidence that the
melting
transition is first order.  This is consistent with our observation that the
2D solid
melts directly into an isotropic liquid. The transition is weakly
first order however, as can be seen from our result that the jump in
$\eta_{\bG_1}(T_m^-)$
(and hence the vortex lattice shear modulus $\mu$) at melting
remains very close to the KTNHY universal value.

%%%%%%%%%%%%%%%%%%%%%%%%%%%%%%%%%%%%%%%%%%%%%%%%%%%%%%%%%%%%%%%%%%%%%%%%%%%%%%%

\subsection{The depinning transition}

Finally, we consider the order of the depinning transition at $T_c(f)$.
In their work on 2D melting on a periodic substrate,
Nelson and Halperin\cite{nhmelt} studied this ``commensurate to floating"
transition using renormalization group techniques. They concluded that the
transition is most likely second order, with properties very similar to the
floating solid to liquid melting transition discussed in Section IIC.
To test this prediction, we use the histogram method applied at the depinning
transition $T_c(f)$, just as we did in the preceding section for melting at
$T_m$.
Measuring the energy
distribution $P(E)$ at $T_c(f)$, for $f=1/49$ and various L, we show the
free energy $F(E)$ versus $E$ in Figure \ref{f9}. As was seen at $T_m$,
we now similarly see a pronounced double
well structure with barrier $\Delta F$ growing with the size of the system
(see the inset). Again, this is a clear indication  that the transition is
first order.
Due to the low acceptance rates at these low temperatures, we had to perform
as many as $4\times 10^7$ MC sweeps for each system size, in order to get
reasonably accurate energy distributions. By finding the temperature that
produces minima of equal depth, we estimate that $T_c(f=1/49)=0.0046$.

\section{Simulations on the square grid: dilute case}

\subsection{The ground state}

In the present section we shall investigate the dilute limit $(f\ll 1)$ of the
Coulomb gas, on a square grid of sites.  This corresponds to a square periodic
superconducting network, which has been the predominant geometry in
experimental
and theoretical studies of networks.\cite{nato,tj,lobb,straley}
Qualitatively, we find similar behavior as found in Section III for the
triangular
grid: a depinning transition $T_c(f)$, from a commensurate pinned solid to
a floating solid, followed at higher temperature by a melting transition $T_m$
to a liquid.

While the case of a square grid is more relevant to the physics of
superconducting arrays, it is somewhat more difficult to study theoretically
than the triangular grid of Section III. The main reason for this is the
rich variety of
ground state configurations that one can encounter for various system sizes and
vortex densities $f$.
The most extensive enumeration of such ground states, for both dilute and dense
$f$, has been carried out by Straley and Barnett.\cite{straley}
This richness in ground state structure is due to the
intrinsic competition between the repulsive
vortex-vortex interaction, which prefers the formation of a  perfect
triangular lattice, and the
geometrical constraints implied by the presence of the square grid.
Since a triangular lattice is incommensurate with a square grid, for small $f$
the resulting ground states form high order commensurate
approximations to a triangular lattice, that vary substantially as $f$ varies.
Thus, while in the triangular grid a density $f=1/m^2$ can always fit
commensurably in a lattice of size $L=s\cdot m$ ($s$ integer), for the square
grid, a density of $f=1/q$ ($q$ integer) will require a lattice of at least
$L=s\cdot q$ to contain the commensurate ground state.  It thus becomes
too difficult to carry out detailed finite size scaling calculations at
small $f$, as the lattice sizes needed quickly become too large to simulate.
We therefore must be content with a more qualitative analysis based on
simulations
at a fixed size system.
A second problem, related to the high order commensurability of the ground
state, is the existence of excited states that are nearly degenerate in energy
with
the ground state.  This can sometimes cause equilibration problems, or leave
uncertainty as to the configuration of the true ground state.
Fortunately, these difficulties occur only at low temperatures, below the
depinning transition $T_c(f)$, where commensurability effects are crucial.
In sufficiently dilute systems, the melting of the floating solid phase at
$T_m$
is largely unaffected by such difficulties, and we find results familiar
to the preceding section.

We have performed simulations for systems with a wide spectrum of densities
$f=1/q$ ($q$ integer) with $10<q<90$.  From inspection of the inverse
dielectric constant $\epsilon^{-1}(T)$ and the orientational correlation
$\varphi_6(T)$, for a fixed size $L$, we estimate  the depinning and melting
transition temperatures,
$T_c(f)$ and $T_m$, and we plot these values versus $f$ in Figure \ref{f10}.
We see that above $f\simeq 1/30$ the depinning  and melting transitions merge,
and there is only a single transition from pinned solid to liquid.
Due to the varying commensurability of the ground state as $f$ varies, the
values $T_c(f)$ and $T_m$ no longer decrease monotonically with $f$, as was
found for
the triangular grid.  Nevertheless, we see that $T_c(f)$ still
tends linearly to zero as $f$ decreases (dashed line), in agreement with the
TJ conjecture.  $T_m$ appears to saturate around $0.007$, in agreement with
the melting temperature found for the triangular grid.

In order to find
the ground state of the system for a given density $f$ and size $L$,
we have devised a simple
program that scans all possible periodic vortex configurations, consistent with
periodic boundary conditions, and evaluates their energy.  When translational
and
inversion symmetries are accounted for, the total
number of distinct configurations is relatively small, and even for the largest
of the systems that we considered it took only few minutes to execute the
program. The
lowest energy configuration obtained in this manner was then taken as a
candidate
for the ground state. In many cases we have verified that this indeed was a
true
ground state by performing a slow MC cooling from a random configuration at
high
temperature. In all cases we found that for $f=1/q$, the
ground state has a $q\times q$ periodicity. In Figure
\ref{f11} we display two typical examples of these ground state configurations.
The almost perfect triangular lattice
$(\sqrt{68}\times\sqrt{68}\times\sqrt{72})$
in Figure \ref{f11}a is for $f=1/60$. Figure \ref{f11}b shows the example
of a nearly square vortex lattice $(\sqrt{50}\times\sqrt{53}\times\sqrt{89})$
with
$f=1/51$. In what follows we shall concentrate on these two special
cases as representatives of two classes of systems with slightly different
physical properties.

\subsection{Systems with ``nearly triangular" ground state: $f=1/60$}

Not very surprisingly, systems with an almost triangular ground state, such as
$f=1/60$ shown in Figure \ref{f11}a, behave in a fashion similar to  systems
on the triangular grid studied in Section III. We display the behavior of the
inverse dielectric function $\epsilon^{-1}(T)$  for $f=1/60$ and $L=60$ in
Figure
\ref{f12}a. In Figure \ref{f12}b we show the six-fold and four-fold
orientational
correlations, $\varphi_6(T)$ and $\varphi_4(T)$.
A sharp drop in $\epsilon^{-1}(T)$ around $T_c(f)\simeq 0.0045$
signals the loss of superconducting phase coherence.  Above $T_c(f)$,
$\epsilon^{-1}$ is zero, but $\varphi_6(T)$ stays finite. Based on our
experience
from the triangular grid, we take this as a signature of a floating triangular
solid with long-range orientational order. $T_c(f)$ is thus a transition
from a commensurate pinned solid, to an incommensurate floating solid.
Around $T_m\simeq 0.0075$ we see
that $\varphi_6(T)$ drops again to very small values; we take this as a signal
that the floating solid has melted into a vortex liquid.

In order to confirm this scenario, we calculate the structure function
$S(\bk)$ at various temperatures, and display the resulting intensity
plots in Figure \ref{f13}. We clearly see the pinned solid
(Figure \ref{f13}a), the floating solid (Figure \ref{f13}b), and the liquid
(Figure \ref{f13}c) phases.
It is interesting to note that the rotational symmetry of the pinned and
floating
solids break the four-fold rotational symmetry of the square grid, leading
to two possible degenerate orientations. In the liquid however, we see that the
four-fold symmetry of the square grid is restored, with a strong four-fold
angular modulation of the circular intensity peaks.
This observation is also confirmed by a direct
measurement of $\varphi_4(T)$ (see Figure \ref{f11}b), which is close to zero
in the floating solid phase, but then rises sharply
at the melting transition and only slowly vanishes with increasing
temperature.   The small values of $\varphi_4(T)$ for $T_c(f)<T<T_m$ are an
indication of the extent to which the commensurate, slightly distorted
triangular
lattice of the ground state, beomes an incommensurate perfect triangular
lattice
in the floating solid phase.

Since we are unable to carry out finite size scaling, we are unable to search
in detail for the hexatic phase, or for the predicted\cite{nhmelt} Ising
transition from  the hexatic to the normal liquid.  However, as the four-fold
symmetry appears to be restored at the same temperature as the melting
transition, we suggest that, as was found for the triangular grid, the
hexatic phase is absent and the melting transition is first order.

Although finite size scaling is not possible, we can nevertheless
still obtain the translational correlation exponent $\eta_{\bG_1}$
by analyzing the decay of the peaks in the structure function, using
the method discussed in connection with
Eqs.(\ref{e21.1})-(\ref{e21.3}). In the present case the implementation
of this method is somewhat trickier than it was for the triangular grid,
since, due to the incommensurability of the floating solid, the
peaks in $S(\bk)$ do not have well defined
positions $\bG$ on the square reciprocal lattice. We
overcome this difficulty by numerically scanning $S(\bk)$ for local maxima at
a given distance from the center of the reciprocal lattice, and averaging
over the
heights
of peaks of the same order. The peak heights estimated in this way are shown in
Figure \ref{f14} for several temperatures $T$ in the floating lattice phase.
Dashed lines are least square fits to the formula (\ref{e21.2}), and the
extracted exponents $\eta_{\bG_1}(T)$ are summarized in Table \ref{t3}. The
accuracy of the fit appears to be as good as in the case of the triangular
grid,
 and
we therefore have good reason to believe that our determination of
$\eta_{\bG_1}(T)$  (and thus the vortex shear modulus $\mu$) is reasonably
accurate.
Once again we see from Table \ref{t3} that $\eta_{\bG_1}(T)$  first exceeds the
universal KT value of $1/3$ at $T\simeq 0.007$, very close to the melting
temperature $T_m\simeq 0.0075$ estimated from the behavior of $\varphi_6(T)$.

%%%%%%%%%%%%%%%%%%%%%%%%%%%%%%%%%%%%%%%%%%%%%%%%%%%%%%%%%%%%%%%%%%%%%%%%%%%%%
\subsection{Systems with ``nearly square" ground state: $f=1/51$}

	We now briefly describe our results from simulations on a system with
density
$f=1/51$, which possesses a nearly square ground state (Figure \ref{f11}b).
Our results are for a system of size $L=51$.
The temperature dependence of the inverse dielectric function
$\epsilon^{-1}(T)$
is shown in Figure \ref{f15}a. From the data, we estimate the depinning
temperature
to be $T_c(f)\simeq 0.0035$.  We note that even though $f=1/51$ here is larger
than the $f=1/60$ studied in the previous section, we find
$0.0035=T_c(1/51)<T_c(1/60)=0.0045$,
thus illustrating the nonmonotonic behavior of $T_c(f)$ for small $f$.
The significantly lower depinning temperature
in the present case may be qualitatively understood as a result of the
larger distortion of the ground state away from the perfect triangular lattice
favored by the vortex-vortex interaction.  This large distortion, which is
favored by the pinning energy, comes at a cost in vortex-vortex interaction
energy.
The result is a reduced free energy difference between the pinned
``distorted triangular" solid and the perfect triangular floating solid, and
hence
a reduced depinning temperature.  A similar observation holds for the other
values
of $f$ we have studied: systems with relatively lower $T_c(f)$ compared to
other nearby values of $f$, are those with greater distortion of the ground
state
from a triangular lattice.

In Figure \ref{f15}b we plot the temperature dependencies of the four-fold
and six-fold orientational
correlations $\varphi_4(T)$ and $\varphi_6(T)$. In contrast to the
previously considered cases, the melting transition is barely visible here:
one sees only a small kink in $\varphi_4(T)$ and an inconspicuous
dip in $\varphi_6(T)$ near $T=0.005$. For a clearer picture of melting, we
show the structure function $S(\bk)$ in Figure \ref{f16}.  We see again
the pinned solid (Figure \ref{f16}a), the floating solid (Figure \ref{f16}b),
and the liquid (Figure \ref{f16}c).
Note that the peaks in the floating solid occur at distinctly different
wavevectors $\bk$ than in the pinned solid; this emphasizes the fact that
$T_c(f)$ is truly a transition from a commensurate ``nearly square'' lattice,
to an incommensurate floating triangular lattice.
Inspection of these intensity plots
gives a melting transition of $T_m\simeq 0.005$, in agreement with the
value hinted at in Figure \ref{f15}b.  We note that this value is significantly
lower than the value of $0.007$ found in other cases.
Thus commensurability effects can also significantly lower the melting
temperature.  Such commensurability effects presumably become less significant
as $f$ decreases and the ground state becomes increasingly closer to a
triangular
lattice.  We thus expect that the melting temperature becomes $T_m\simeq 0.007$
in the asymptotic $f\to 0$ limit, as is indeed suggested by Figure \ref{f10}.
\section{Simulations on the square grid: near full frustration}

The square superconducting network with $f=1/2$ has been the focus
of extensive theoretical study in recent years \cite{half,jrl1,grest}.
As the Hamiltonians, Eq.(\ref{eqH}) and Eq.(\ref{e3}), are periodic in
$f$ with period $1$, $f=1/2$ represents the strongest magnetic field,
and most dense vortex configuration, discernable by the network.
Thus this case is usually refered to as ``fully frustrated."
The ground state of this configuration is in some sense the simplest of
all $f>0$, consisting of a checkerboard pattern of vortices,
with $n_i=1$ and $n_i=0$ on the two alternating sublattices of the square grid.
This dense vortex lattice melts\cite{jrl1} directly into a vortex liquid at
$T_m(1/2)\simeq 0.13$.  Superconducting coherence vanishes\cite{note12}
at $T_c(1/2)\simeq T_m(1/2)$.

We now wish to study how the system behaves as $f$ is varied slightly away
from $1/2$, in order to test the discontinuous behavior predicted by the
TJ conjecture.  We study in particular systems with $f=1/2-1/q$, with
integer $q$ large.  While the ground states for densities of this form
have been studied by Straley and co-workers,\cite{straley,kolachi}
finite temperature properties have remain unexplored.

%%%%%%%%%%%%%%%%%%%%%%%%%%%%%%%%%%%%%%%%%%%%%%%%%%%%%%%%%%%%%%%%%%%%%%%%%%%%%%
\subsection{$f=5/11$}

We first consider the particular case of $f=5/11$, which may be written as
$f=1/2-1/22$.  The correct ground state for this case, which we show in
Figure \ref{f17}, was first found by
Kolachi and Straley.\cite{kolachi}  It consists of a periodic
superlattice of vortex vacancies superimposed on an otherwise uniform $f=1/2$
like
background, and is periodic with a $22\times 22$ unit cell.
Our motivation is to see whether or not this superlattice of vacancies
(or ``defects")
can melt independently of the $f=1/2$ like background, and if so, whether
the resulting liquid of vacancies destroys superconducting coherence.
Our analysis is similar to that in the previous section.

Heating from the ground state, we show sample intensity plots of the
structure function $S(\bk)$, at different temperatures, in Figure \ref{f18}.
Figure \ref{f18}a shows the low temperature phase at $T=0.010$.
The bright Bragg peaks at $\bk =(\pm\pi/a_0,\pm\pi/a_0)$ originate from the
vortex ordering in the $f=1/2$ like background, while the periodic Bravais
lattice of less intense Bragg peaks is due to the
defect superlattice, which at this low temperature is pinned to the substrate.
Thus we have a ``pinned defect solid" phase.  As the temperature is increased,
we find that the defect superlattice melts at $T_m\simeq 0.015$.  In
Figure \ref{f18}b we show the system at $T=0.018$, just above this melting.
The defects no longer give rise to Bragg peaks, but instead we see the
circular rings (with strong four-fold angular modulation) characteristic of
a defect liquid.  However the bright Bragg peaks at $\bk =
(\pm\pi/a_0,\pm\pi/a_0)$
remain, indicating that the $f=1/2$ like vortex background remains ordered.
Upon increasing the temperature further, the
ordered $f=1/2$ background is also eventually destroyed at $T_{m^\prime}
\simeq 0.040$.
In Figure \ref{f18}c we show the system at $T=0.055$, above $T_{m^\prime}$.
The peaks at $\bk =(\pm\pi/a_0,\pm\pi/a_0)$ have broadened to finite width,
indicating the disordering of the $f=1/2$ like background.

To see the melting transitions more clearly, in Figure \ref{f19}
we plot versus temperature the peak heights $S({\bf q^*})$ and $S(\bG_1)$, with
$\bq^* \equiv(\pi/a_0,\pi/a_0)$ giving the ordering of the $f=1/2$
like background, and
$\bG_1$ the shortest reciprocal lattice vector of the defect superlattice.
We see that $S(\bG_1)$ vanishes sharply at $T_m\simeq 0.015$, where the
pinned defect superlattice melts into a defect liquid.
$S({\bf q}^*)$, however, remains at its $T=0$ value of unity for all
temperatures
up to $T\simeq 0.020$, clearly demonstrating that the $f=1/2$
like vortex background remains ordered throughout the defect melting
transition.
$S({\bf q}^*)$ starts to drop to zero around $T_{m^\prime}\simeq 0.04$,
where, based on
the structure function intensity plots, we have estimated that the $f=1/2$
like background melts.

To investigate superconducting coherence,
in Figure \ref{f20} we show the inverse dielectric function versus temperature.
We see that $\epsilon^{-1}$ vanishes at the defect melting transition $T_m$.
The diffusing defects above $T_m$ induce a diffusion of vortices, which
must move to fill in the ``hole" left behind by the defect as it moves.
The diffusing vortices are then responsible for the destruction of
superconducting phase coherence.

To summarize, we have found clear evidence that the introduction of a small
concentration of defects into the fully frustrated system results in a
dramatic decrease of the superconducting transition temperature from its
$f=1/2$ value.  The fluctuations of the defect superlattice, on an
essentially frozen $f=1/2$ like background, result in behavior which is
in many respects like that of the dilute vortex lattices studied in Section IV.
In the present case, we find that the defect superlattice melts directly from
a pinned solid into a liquid.  In the following section we will argue,
following the analogy with Section IV, that a more dilute defect superlattice
would first unpin at a $T_c(f)$ into a floating defect superlattice, which
would then melt at a higher temperature $T_m$ into a defect liquid.

The sharp melting transition of the $f=1/2$ like background at a
temperature $T_{m^\prime}$ distinctly higher than the defect melting at $T_m$
is a new phenomenon, with no analogue in the dilute small $f$
systems (at small $f$ there is only a smooth crossover remnant of
the vortex-antivortex unbinding transition of the $f=0$ case).
 From symmetry, one would expect that the transition at $T_{m^\prime}$
is of the Ising type.  Undersanding whether the melting transition in the
pure $f=1/2$
case is Ising like or not, has been the subject of much work, with the
most recent simulations suggesting that it is not;\cite{jrl1} if it is not
Ising, this is most likely due to the long range nature of the vortex
interactions.
For the $f=5/11$ case however, the melted defect liquid will serve to screen
the interactions of the vortices in the $f=1/2$ like background, resulting
in effectively short ranged interactions.  An Ising transition is therefore
most probable.  We are unable to test this prediction, as we are unable
to carry out a detailed finite size scaling analysis, for the same reasons
as discussed in Section IV.  However
the strong screening effect of the defect liquid is evident in
the substantial reduction of  the background melting temperature,
$T_{m^\prime}(5/11)\simeq 0.04$, as compared to the melting transition
$T_m(1/2)\simeq 0.13$ of the pure $f=1/2$ case.

%%%%%%%%%%%%%%%%%%%%%%%%%%%%%%%%%%%%%%%%%%%%%%%%%%%%%%%%%%%%%%%%%%%%%%%%%%%%%
\subsection{General case: $f=1/2-1/q$}

In this section, we strengthen the analogy between the melting of the
defect superlattice seen in the previous section, with the melting of
the dilute (small $f$) vortex lattices studied in Section IV, in order
to discuss the general case of $f=1/2-1/q$.  Our goal is to establish that
for a more dilute density of defects, one will have a floating defect solid
intermediate between a pinned defect solid and a defect liquid.

As seen for the $f=5/11$ case, throughout a temperature range including
the defect lattice melting transition $T_m$, the $f=1/2$ like background
vortices remain perfectly ordered as at $T=0$; domain excitations, which
would reduce $S({\bf q}^*)$ from its $T=0$ value of unity, become important
only above $T_m$.  In this case, one can focus solely on excitations
which are due to the motion of the defects.  At $T=0$, these defects are
seen to sit on the same sublattice of the square grid as do the $n_i=1$
vortices of the $f=1/2$ like background (see Figure \ref{f17});
equivalently, one never has two vortices on two nearest neighbor sites.
We assume that this restriction continues to hold at finite temperatures
up to and including $T_m$, i.e. the cost in energy to have two vortices
on nearest neighbor sites is so high compared to $T_m$, that such excitations
may be ignored.

With this assumption, we have reduced our problem at
$f=1/2-1/q$ to that of
a density $1/q$ of logarithmically interacting defects, which are restricted
to move on only one of the two sublattices of the original square grid.
This sublattice is itself a square lattice of lattice constant $\sqrt 2 a_0$.
As the sublattice has
half the number of sites as in the original grid, our problem is thus
effectively the same as a dilute density $f^\prime=2/q$ of vortices
on a square grid.  This mapping would be exact (within our assumptions)
except for the fact that the interaction potential $V(\br)$ between the
defects is still defined with respect to the original square grid, and not
the sublattice to which the defects are constrained.  However, as $q$ gets
large, and the average spacing between defects becomes much greater than
$a_0$, we expect that this difference will be a negligible effect.

The assumption that the defects move only on one sublattice can be checked
for the $f=5/11$ case of the previous section.  Restricting the defects
in real space to a sublattice whose unit cell has twice the area of
that of the orginial square grid, means that
the 1st Brillouin Zone of the effective reciprocal lattice is reduced by a
factor of one half.  Instead of the square shaped BZ shown in Figures
\ref{f18},
the  effective BZ is now an inscribed diamond whose vertices bisect
the edges of the squares of
Figures \ref{f18}.  The structure function $S(\bk)$, as plotted over the
full square shaped BZ of the original square grid, should now just be obtained
by
a periodic repetition of the diamond shaped BZ corresponding to the sublattice.
Such periodicity is clearly seen in Figures \ref{f18}a and \ref{f18}b, for
both the
pinned defect solid, and the melted defect liquid.  It
is absent in Figure \ref{f18}c, where $T>T_{m^\prime}$, and the $f=1/2$
like background has melted.

Having checked the validity of our assumption for $f=5/11$, we note that
that it should be even better satisfied for more dilute defect densities
$f=1/2-1/q$,
$q>22$.  As $q$ increases, the density of defects decreases, resulting
a reduced screening of the interactions between the background vortices.
The background melting temperature $T_{m^\prime}$ should therefore increase and
approach its higher $f=1/2$ value.  At the same time, the defect superlattice
unpinning temperature $T_c$ should decrease as $\sim 1/q$, while the
defect superlattice melting temperature $T_m$ saturates to a lower fixed value.
Thus we expect that the window of temperatures in which our assumption is valid
becomes wider as $q$ increases.

We can now understand the behavior found in the preceeding section.
For $f=5/11=1/2-1/22$, we have
$q=22$, and so the defects behave like an effective vortex density of
$f^\prime=2/q=1/11$.  Comparing to our results of Figure \ref{f10} in
Section IV, we see that $f^\prime$ is large enough that we expect only a
direct melting of the pinned defect solid to a defect liquid, consistent
with our observation in the preceeding section.  In order to observe a floating
defect solid, we will have to consider an $f^\prime<1/30$, or an $f=1/2-1/q$
with $q>60$.

To simulate a system with $f=1/2-1/60$ directly, would require a grid size
of at least $L=60$, with $N_c=1740$ vortices.  This is beyond our present
computational ability ($f=5/11$, with $L=22$ and $N_c=220$, is about the
largest system we can manage).  However our conclusion,
that at temperatures low compared to $T_{m^\prime}$ the defects move
in the presence of an effectively frozen $f=1/2$ like background,
allows us to construct a much more efficient algorithm which will be suitable
for describing behavior up to and above the defect melting transition $T_m$,
provided we stay below the background melting transition $T_{m^\prime}$.
We do this by fixing the $f=1/2$ like
background and allowing only the vacancies to move around.
This significantly reduces the number of degrees of freedom in the
simulation, and we shall thus be able to treat systems with a much smaller
fraction
of defects $1/q$, than we could by direct MC simulation.

In order to implement the algorithm suggested above, we formally decompose the
charge at site $\br_i$ into two parts
\begin{equation}
n_i=s_i-\delta n_i,
\label{e30}
\end{equation}
where
\begin{equation}
s_i\equiv {1\over 2}[1+(-1)^{x_i+y_i}] = {1\over 2}(1+e^{i\br_i\cdot\bq^*})
\label{e31}
\end{equation}
is the staggered pattern of the background vortices
($\bq^*\equiv (\pi/a_0,\pi/a_0)$) and
$\delta n_i$ are new integer variables representing the defects in the
background.  Neutrality requires that $\sum_i\delta n_i=L^2/q$.
Substituting Eq.(\ref{e30}) for $n_i$ in the Hamiltonian
(\ref{e3}), we get
\begin{equation}
{\cal H}={1\over 2}\sum_{ij} \dn_i V'_{ij}\dn_j - \sum_{ij} \dn_i
V'_{ij}(s_j-f)
         +{1\over 2}\sum_{ij}(s_i-f) V'_{ij}(s_j-f),
\label{e32}
\end{equation}
where $V'_{ij}\equiv V'(\br_i-\br_j)$. The first term in the
Hamiltonian (\ref{e32}) gives the interaction between defects; the second term
represents the interaction of the defects with a one-body potential
\begin{equation}
\Phi_i\equiv\sum_j V'_{ij}(s_j-f)
\label{e33}
\end{equation}
created by the background; the last term is just an additive constant.
Substituting Eq.(\ref{e31}) for the $s_i$ into Eq.(\ref{e33}) above,
gives the potential $\Phi_i$ in terms of the Fourier components of
the interaction, $V_\bk$,
\begin{equation}
\Phi_i={1\over 2} V_{\bq^*} e^{i\br_i\cdot\bq^*} - ({1\over 2}-f)
\sum_{\bk\neq 0} V_\bk,
\label{e34}
\end{equation}
where from Eq.(\ref{e6}) we have $V_{\bq^*}=\pi/4$.  Thus $\Phi_i$ oscillates
with the same checkerboard pattern of the $f=1/2$ like background.
Comparing with Eq.(\ref{e31}), we see that
the Hamiltonian of Eq.(\ref{e32}) can now be rewritten in the following
simple form
\begin{equation}
{\cal H}={1\over 2}\sum_{ij} \dn_i V'_{ij}\dn_j -
         {\pi\over 4}\sum_i\dn_i  s_i + E_0,
\label{e35}
\end{equation}
where $E_0$ is an additive constant.

So far, the formulation above is exact.  Our approximation that the background
is frozen, and that defects only move on the sublattice defined by $s_i=1$,
occurs when we consider only the case where $N_c$ sites have the value
$\dn_i=1$, and all other sites have $\dn_i=0$.
In this approximation, the Coulomb gas near full frustration, $f=1/2-1/q$,
is equivalent at low temperatures to the dilute Coulomb
gas of defects with integer charges $\dn_i=0,1$, moving in a staggered
potential
of magnitude $\delta\Phi=\pi/4=0.7853\dots$.
As this magnitude is about two orders of magnitude greater than the relevant
excitation energy scale, set by temperature $T_m$, the sites with $\delta
n_i=1$
are essentially restricted to the sublattice where $s_i=1$; in this case
they represent the vacancies in the $f=1/2$ like background.  The case where
$\dn_i=1$ on the opposite sublattice where $s_i=0$, represents a
$(+1,-1)$ vortex-antivortex excitation, which can be ignored on energtic
grounds
as we had shown in earlier sections.

To check the consistency of the above procedure,
we have redone our simulation of $f=5/11$ using
the new algorithm based on the Hamiltonian (\ref{e35}). In a fraction of the
CPU time
needed for the original simulation using the full Hamiltonian (\ref{e3}),
we have
recovered our original results for all quantities, at all temperatures up
to about
$T\simeq 0.040$, where fluctuations in the $f=1/2$ background become
significant.

Having verified the consistency of the new algorithm in this way,
we now proceed to simulate systems with more dilute concentrations of defects.
In Figure \ref{f21}, we display the structure function $S(\bk)$ for the case
$f=22/45=1/2-1/90$. As expected from the discussion above, we observe a clear
signature of the floating solid phase (Figure \ref{f21}b) in the temperature
range $0.005<T<0.008$. This range is identical to the range
in which we found the floating solid phase for $f^\prime=1/45$ (see Figure
\ref{f10}).
The low temperature phase is a familiar ``pinned defect solid'' (Figure
\ref{f21}a); the high temperature phase is a defect liquid with strong
four-fold
correlations (Figure \ref{f21}c). The above scenario is confirmed by a direct
measurement of $\epsilon^{-1}(T)$ and the six-fold orientational correlation
$\varphi_6(T)$
of the defects $\dn_i$, shown in Figure \ref{f22}.
Both quantities behave in a way similar
to those measured for dilute vortex systems, showing a sharp drop in
$\epsilon^{-1}(T)$ at the depinning transition, and a plateau in $\varphi_6(T)$
in the floating phase.

To summarize, we conclude that for $f=1/2-1/q$, with $q>60$, there will be
the following sequence of transitions.  At low temperature there is a
pinned superlattice of defects of density $1/q$,
which unpins at $T_c(f)\sim 2/q$ into
a floating superlattice of defects.  This floating lattice melts at
$T_m\simeq 0.007$ into an isotropic defect liquid.  Finally, at
$T_{m^\prime}$, which approaches the value of 0.13 as $q$ increases,
the $f=1/2$ like background melts via an Ising transition, resulting in
an isotropic vortex liquid of density $f$.

\section{Summary and conclusions}

We have carried out extensive Monte Carlo simulations of the Coulomb gas
Hamiltonian (\ref{e3}) as a model of a 2D superconducting network in
an external transverse magnetic field.
One of the goals
of our work was to systematically study, for the special cases of vortex
density
$f=1/q$ and $f=1/2-1/q$ ($q\gg 2$) a conjecture put forward by Teitel and
Jayaprakash,\cite{tj} that for $f=p/q$ the superconducting transition
temperature
scales approximately as $T_c(f)\sim 1/q$.  For the dilute case, $f=1/q$, we
have found good agreement with this
conjecture, provided one interprets the superconducting transition temperature
to be the vortex lattice unpinning temperature $T_c(f)$,
where the ground state vortex
lattice decouples from the superconducting network, and is free to slide
transversly to any applied d.c.\ current, thus producing ``flux flow"
resistance.

A new result of our work is the realization that above $T_c(f)$, for
sufficiently
dilute systems, a depinned ``floating" vortex solid will exist.  This floating
vortex solid has essentially the same properties as a vortex lattice in a
uniform
superconducting film, and it melts (as $q\to\infty$) at $T_m\simeq 0.007$
into an isotropic vortex liquid.  While the true onset of finite linear d.c.\
resistivity will be $T_c(f)$, the melting at $T_m$ is presumably accompanied
by a sharp rise in resistivity.  The distinction between $T_c$ and $T_m$,
however,
may be difficult to observe experimentally, due to the
existence of large energy barriers\cite{barrier} for the hopping
of a vortex between neighboring cells of the superconducting network.
As discussed in the introduction, the discrete nature of the network introduces
an effective periodic pinning potential for vortices.  For a square Josephson
array, Lobb {\it et al.}\cite{barrier} have estimated the energy barrier of
this pinning potential to be  $E_{b}\simeq 0.199/2\pi=0.0317$
(in our energy units\cite{notebar}).
This is almost five times the vortex lattice melting temperature $T_m\simeq
0.007$!
Thus for the square network, one is most likely to observe upon cooling only
a vortex liquid, in which the vortex mobility decreases exponentially as
$e^{-E_{b}/T}$; the true phase transitions at $T_m$ and $T_c$ will be masked
by the extremely slow relaxation over the energy barriers $E_{b}$ at these low
temperatures.
Such behavior has in fact been reported\cite{expt1,expt2}
in experimental studies of square Josephson
arrays, where for small $f$ near $f=0$, only exponentially decreasing resistive
tails
are observed at low temperature; no evidence for the melting or depinning
transitions at $T_m$ and $T_c$ has been found.
For the triangular Josephson array however (a case we have not explicitly
studied here), the energy barrier
is estimated \cite{barrier} to be $E_{b}\simeq 0.0427/2\pi=0.0068$ (in our
energy
units).
This is
comparable to $T_m$, and so there might be some slight chance of experimentally
observing the melting transition.
Recent experimental studies of this system\cite{expt3}
at small $f$ have found surprising
dynamical behavior, indicating anomalously slow diffusion of vortices.
However the temperature of these experiments, $T\simeq 0.5 T_c(f=0)$,
appears to be too high for these results to be explained by any of the
melting or depinning effects we have found here.
For a honeycomb Josephson array (vortices
on a triangular grid) we estimate the highest barrier,\cite{notebar2}
 $E_{b}\simeq 0.751/2\pi = 0.119$.

For dense systems close to full frustration, $f=1/2-1/q$, we have argued that,
at temperatures low compared to the melting temperature of the pure $f=1/2$
system,
the physics is dominated by defects moving on top of a quenched $f=1/2$ like
vortex
background.  We have shown how this system of defects can be mapped onto the
dilute Coulomb gas of vortices with density $f^\prime=2/q$.  The resulting
behavior is then obtained from knowledge of this dilute limit.  The TJ
conjecture again holds, with $T_c(f=(q-2)/2q)\sim 2/q$ marking the transition
from a pinned defect superlattice to a floating defect superlattice, in which
true d.c. superconductivity is lost.  At a higher $T_m\simeq 0.007$, the
floating defect solid melts into an isotropic defect liquid.  At yet a higher
$T_{m^\prime}$, the $f=1/2$ like background disorders.  As in the $f=0$ case,
the transitions at $T_c$ and $T_m$ may be difficult to observe experimentally,
due to the energy barrier for a defect to hop between nearest neighbor
sites of the relevant sublattice.  For a square lattice, Dang and
Gy\"orffy\cite{dang} have estimated
this barrier to be $E_b\simeq 0.368/2\pi$ (in our energy units), even
larger than that found for $f=0$.
Experimental studies\cite{expt1,expt33} of square Josephson arrays for $f$
near $f=1/2$ have again found only exponential resistive tails as the
temperature decreases.

In contrast to the transitions at $T_c$ and $T_m$,
we expect that the sharp disordering transition
of the $f=1/2$ like background at $T_{m^\prime}$ should be
experimentally observable.  This follows since
for $f=1/2-1/q$, we expect that as $q\to\infty$, $T_{m^\prime}\to T_m(1/2)
\simeq 0.13$, well above the energy scale of the barriers.  This
transition would presumably manifest itself as a singular increase
in the linear resistivity at $T_{m^\prime}$ (from an already finite
value).  The phase boundary $T_{m^\prime}(f)$ near $f=1/2$ is presumably
a smooth funtion of the defect density, $1/2-f$, however we
are unable to estimate it due to our inability to simulate sufficiently
large systems.

Our mapping between
a dilute density of defects near $f=1/2$, and a dilute vortex density near
$f=0$, may be extended to the more general case.  Using the same arguements
as in Section V, we would expect that the system with $f=1/2-p/q$, with
$p/q$ sufficiently small, should have the same low temperature behavior
as the density $f^\prime=2p/q$.  For $p/q$ sufficiently small, we would expect
that the dilute vortex lattice of density $f^\prime=2p/q$ behaves qualitatively
like those of density $1/q$ studied here, i.e. there is first a depinning
transition at
$T_c(f)$ which decreases as $p/q$ decreases (whether it vanishes as $p/q$ or
as $1/q$ remains to be investigated) followed by a melting transition at
$T_m\simeq 0.007$.  Thus for any rational fraction sufficiently close to either
$f=0$ or $f=1/2$, we would expect behavior similar to the cases explicitly
studied here.

We thus see that the TJ conjecture appears to hold, according to the
following scenario.  Consider a vortex density $f$ close to some simple
fraction $f_0=p_0/q_0$, $f=f_0-1/q$, with $q_0 \ll q$.
The ground state is one which is almost everywhere like
that of $f_0$, except for a pinned periodic superlattice of defects of density
$1/q$.  If $q$ is sufficiently large, this superlattice will unpin into
a floating defect solid at $T_c(f)\sim1/q$.
Defects which are free to move lead to flux flow resistance, and
destroy the superconducting phase coherence
of the system. Thus an arbitrarily small concentration of defects added on top
of the $f_0$ like background (i.e. for $f$ arbitrarily close to $f_0$)
dramatically decreases the superconducting transition
temperature, when compared to pure $f_0$ system.
We have explicitly tested this scenario for the cases $f_0=0$ and $f_0=1/2$.
We speculate that this
behavior will be characteristic of  systems near any simple fraction
$f_0=p_0/q_0$.  We further speculate that his behavior
may be characteristic for $any$ sufficiently small rational fraction of defects
away from a simple fraction $f_0$, i.e. $f=p_0/q_0-p/q$ with $q_0\ll
q$.

A second goal of our work was to study in detail the melting transition of
the 2D vortex lattice. This problem has been
addressed previously only in the context of uniform superconducting films.
Here we have addressed this issue in the context of a superconducting network.
We believe, however, that our results for the dilute case we have studied in
Section III are representative of the continuum limit, as treated within
the London approximation.  Melting within this London approximation has been
treated theoretically by Huberman and Doniach,\cite{hd} and
Fisher\cite{fisher},
who applied the KTNHY theory of defect mediated melting in 2D.  This theory
predicts a second order melting transition at a $T_m$ well below the
Ginzburg-Landau mean field transition temperature $T_{MF}$, as well
as an intermediate
hexatic liquid phase.  We have carried out the first detailed finite size
scaling analysis to check this KTNHY theory as applied to vortex lattice
melting.
We find a value $T_m\simeq0.007\pm0.0005$ in good agreement with the value
estimated by Fisher.  We also find that the vortex lattice
shear modulus jumps discontinuously to zero at $T_m$, with a value close to
the KTNHY prediction.  However we find that the melting transition is weakly
first order, and we find no evidence for a hexatic phase.  Our value for $T_m$,
and our conclusion concerning the order of the melting transition,
are in agreement with earlier simulations\cite{plasma}
of the continuum one component plasma model of Eq.(\ref{eqHCGc}).

This problem of 2D vortex lattice melting has been the focus of much renewed
work recently, due to its potential connection with behavior in anisotropic
high temperature superconductors.
The very existence of a vortex lattice at any finite temperature has been
challenged
by Moore,\cite{moore} who argued
that fluctuations in the phase of the order parameter $\psi(\br)$, due
to shear excitations of the vortex lattice, will cause
the order parameter correlation function $\langle\psi^*(\br)\psi(0)\rangle$
to decay exponentially at any finite temperature.  From such decay, Moore
argued
first for the absence of a superconducting state, and then concluded as a
result of this absence of superconductivity, that
the vortex lattice should not exist.  Support for
this scenario is suggested by high temperature perturbative
expansions, which also find no evidence for freezing into a vortex lattice,
even when evaluated to high order. \cite{expand}

Recently, simulations have been carried out to address this question.  In
contrast
to our work in the London approximation, these works have been carried out
in the so called lowest Landau level (LLL) approximation, which is based
on the Ginzburg-Landau (GL) free energy ${\cal H}_{GL}$ of Eq.(\ref{eqF}).
In this approximation, the complex
order parameter $\psi(\br)$ is expanded in terms of the lowest
degenerate eigenstates of the Gaussian
part of the Ginsburg-Landau free energy,
and the coefficients of this expansion (or
alternatively the complex positions of the vortices) are used as fluctuating
variables in a Monte Carlo simulation with ${\cal H}_{GL}$ as the Hamiltonian.
Using such simulations, and modeling a 2D system by the surface of a sphere,
O'Neill and Moore \cite{oneill} failed to find evidence for a vortex lattice.
Other simulations in a 2D plane however,\cite{LLL1,LLL2,LLL3,sasik}
reported clear evidence for the
melting of a vortex lattice at a finite temperature.  Hu and
MacDonald\cite{LLL2} and Kato and Nagaosa\cite{LLL3}
find that this melting transition is weakly first order, in
agreement with our London result.  \v{S}\'{a}\v{s}ik and  Stroud \cite{sasik}
similarly find a first order transition; they further compute the vortex
lattice
shear modulus $\mu$ and find behavior at $T_m$ in agreement with our
result.
Most recently, Herbut  and Te\v{s}anovi\'{c}\cite{tesanovic}
have developed a density functional
theory of the vortex lattice melting transition, based on the LLL formalism.
They again find results consistent with the above, for the order of the
transition, and the shear modulus $\mu$.

Thus, with the exception of Ref.\onlinecite{oneill}, results from
the London and LLL approximations seem to be in agreement.  This is as
one might expect from the principle of universality in phase transitions.
Although the London approximation at the ``microscopic" level ignores
fluctuations
in the amplitude of $\psi(\br)$ (such as are included in the LLL formalism),
upon coarse graining the London model, phase fluctuations at the
microscopic length scale will generate amplitude
fluctuations on the coarse grained length scale.  On this coarse grained scale,
the system will be described by some effective Ginzburg-Landau  free energy,
complete with amplitude fluctuations,
although higher order terms in $\psi$ beyond those given in Eq.(\ref{eqF})
may be present.
In contrast to the London approximation, the Ginzburg-Landau free energy of
Eq.(\ref{eqF}) includes amplitude fluctuatons at the ``microscopic" scale,
and it is thus often viewed as a more fundamental description.  However,
the GL form of Eq.(\ref{eqF}) represents only the lowest terms in an
expansion of the free energy
in powers of $|\psi|$, and hence is valid only near the mean field
transition $T_{MF}$ where $|\psi|$ is small.  Since vortex lattice melting
occurs
at $T_m<<T_{MF}$, higher order terms in $\psi$ may well be important for
a quantitative description.  These higher order terms, however, are presumably
irrelevent in determining the critical behavior, thus leading to agreement
between the London and LLL simulations.

\section*{Acknowledgments}

The authors are grateful to T. Chen, D. A. Huse, D. R. Nelson, and
Z. Te\v{s}anovi\'{c} for helpful discussions.
This work was supported by DOE grant DE-FG02-89ER14017.  One of us (M.F.)
acknowledges the Rush Rhees Fellowship of the University of Rochester for
additional support.

\bibliographystyle{unsrt}

\begin{thebibliography}{99}

\bibitem{nato} For a review see {\it Proceedings of the NATO Advanced
	Research Workshop on Coherence in Superconducting Networks, Delft,
1987},edited by J. E. Mooij, and G. B. J. Sch\"on, Physica {\bf B142},
      1-302 (1988).
\bibitem{ffh} D. S. Fisher, M. P. A. Fisher, and D. A. Huse, \prb{\bf  43},
	130 (1991).
\bibitem{gk} L. I. Glazman and A. E. Koshelev, \prb{\bf 43}, 2835 (1991).
\bibitem{moore} M. A. Moore, \prb{\bf 39},136 (1989); ibid. {\bf 45}, 7336
(1992).
\bibitem{tj} S. Teitel and C. Jayaprakash, \prl{\bf 51}, 1999 (1983).
\bibitem{lobb} C. J. Lobb, Physica {\bf B126}, 319 (1984).
\bibitem{straley} J. P. Straley and G. M. Barnett, \prb{\bf 48}, 3309 (1993).
\bibitem{kosterlitz} W. Y. Shih and D. Stroud, \prb{\bf 30}, 6774 (1984);
         Y.-H. Li and S. Teitel, \prl{\bf 65}, 2595 (1990);
         Y.-H. Li and S. Teitel, \prl{\bf 67}, 2894 (1991).
\bibitem{half} S. Teitel and C. Jayaprakash, \prb{\bf 27}, 598 (1983);
               P. Minnhagen, \prb{\bf 32}, 7548 (1985);
               D .B. Nicolaides, J. Phys. A {\bf 24}, L231 (1991);
               J. Lee, J. M. Kosterlitz, and E. Granato, \prb{\bf 43}, 11 531
(1991);
               G. Ramirez-Santiago and J. V. Jos\'{e}, \prl{\bf 68}, 1224
(1992); \prb {\bf 49}, 9567 (1994).
\bibitem{jrl1} J.-R. Lee, \prb{\bf 49}, 3317 (1994).
\bibitem{lee} J.-R. Lee and S. Teitel, \prb{\bf 46}, 3247 (1992).
\bibitem{martinoli} For a recent example see, R. Th\'{e}ron, S. E. Korshunov,
    J. B. Simond, Ch. Leemann, and P. Martinoli, \prl{\bf 72}, 562 (1994).
\bibitem{halsey} T. C. Halsey, \prl{\bf 55}, 1018 (1985);
      Physica {\bf B152}, 22 (1988).
\bibitem{barrier} C. J. Lobb, D. W. Abraham, and M. Tinkham, \prb{\bf 27}, 150
   (1983).
\bibitem{expt1} M. S. Rzchowski, S. P. Benz, M. Tinkham, and C. J. Lobb,
    \prb{\bf 42}, 2041 (1990).
\bibitem{straley2} J. P. Straley, \prb{\bf 38}, 11225 (1988).
\bibitem{beck} A. Vallat and H. Beck, \prl{\bf 68}, 3096 (1992).
\bibitem{ktmelt} J. M. Kosterlitz and D. J. Thouless, J. Phys. C {\bf 6},
	1181 (1973).
\bibitem{nhmelt} B. I. Halperin and D. R. Nelson, \prl{\bf 41}, 121 (1978);
     ibid. {\bf 41}, 519 (1978); D. R. Nelson and B. I. Halperin,
     \prb{\bf 19}, 2457 (1979).
\bibitem{young} A. P. Young, \prb{\bf 19}, 1855 (1979).
\bibitem{grest44} G. S. Grest, P. M. Chaikin, and D. Levine, \prl {\bf 60},
1162
(1988)
\bibitem{villain} J. Villain, J. Phys. (Paris) {\bf 36}, 581 (1975).
\bibitem{jkkn} J. V. Jos\'{e}, L. P. Kadanoff, S. Kirkpatrick, and
    D. R. Nelson, \prb{\bf 16}, 1217 (1977);
    E. Fradkin, B. Huberman, and S. Shenker, \prb{\bf 18}, 4789 (1978).
\bibitem{minnrmp} P. Minnhagen, Rev. Mod. Phys. {\bf 59}, 1001 (1987).
\bibitem{note} While we expect that the possible thermodynamic states
 for the cosine
    and the Villain interactions will be the
    same, there has been some suggestion that there may be non-universal
    behavior in such models, that can result in differing critical
    exponents for differing interaction potentials; see the work by
    J. Lee {\it et al}. in Ref.\ \onlinecite{half}.
\bibitem{ohta} T. Ohta and D. Jasnow, \prb{\bf 20}, 130 (1979).
\bibitem{pearl} J. Pearl, Appl. Phys. Lett. {\bf 5}, 65 (1964).
\bibitem{tinkham} M. Tinkham, {\it Introduction to Superconductivity},
    (R.E. Krieger Co., Malabar, FL, 1980).
\bibitem{nhsuper} B. I. Halperin and D. R. Nelson, J. Low Temp. Phys.
    {\bf 36}, 1165 (1979).
\bibitem{hd} S. Doniach and B. A. Huberman, \prl{\bf 42}, 1169 (1979);
     B. A. Huberman and S. Doniach, \prl{\bf 43}, 950 (1979).
\bibitem{fisher} D. S. Fisher, \prb{\bf 22}, 1190 (1980).
\bibitem{mermin} N. D. Mermin and H. Wagner, \prl{\bf 17}, 1133 (1966).
\bibitem{LLtext} L. D.  Landau and E. M. Lifshitz, {\it Theory of Elasticity},
(Pergamon, New York, 1970).
\bibitem{nohex} For a  review  see D. R. Nelson,
  Defect-mediated Phase Transitions, in {\it Phase Transition and Critical
Phenomena},
  Vol. 7, edt. by C. Domb and J. L. Lebowitz (Academic press 1983).
\bibitem{grest} G. S. Grest, \prb{\bf 39}, 9267 (1989).
\bibitem{FrTe} M. Franz and S. Teitel, \prl{\bf  73}, 480 (1994).
\bibitem{hist} J. Lee and J. M. Kosterlitz, \prl{\bf 65}, 137 (1990).
\bibitem{extrapolate} A. M. Ferrenberg and R. H. Swendsen, \prl{\bf 61},
2635 (1988).
\bibitem{note12} Whether $T_c$ is equal to, or slightly lower than, $T_m$ for
  the $f=1/2$ model, remains a question of some controversy.  See
  Refs. \onlinecite{half,jrl1}.
\bibitem{kolachi} M. R. Kolachi and J. P. Straley, \prb{\bf 43}, 7651 (1991).
  See also S. Teitel, Physica B {\bf 152}, 30 (1988), and T. C. Halsey, \prb
{\bf 31}, 5728 (1985), for earlier attempts to find this ground state.
\bibitem{notebar} Due to our rescaling of the Coulomb gas temperature by
  a factor $2\pi J_0$ (see discussion following Eq.(\ref{e4})), the energy
barrier
  in our units is a factor $2\pi$ smaller than the value quoted in
  Ref. \onlinecite{barrier} for a Josephson array.
\bibitem{expt2} H. S. J. van der Zant, H. A. Rijken, and J. E. Mooij,
    J. Low Temp. Phys. {\bf 79}, 289 (1990).
\bibitem{expt3} R. Theron, J.-B. Simond, Ch. Leemann, H. Beck, and P.
Martinoli,
   \prl{\bf 71}, 1246 (1993).
\bibitem{notebar2} Our estimate for the barrier in the honeycomb network
follows
   the calcuation presented in Ref. \onlinecite{barrier}.
   Note that we have cited here energy barriers as computed for Josephson
   arrays with a cosine interaction between superconducting nodes, while we
have
   made comparison to transition temperatures $T_c$ and $T_m$ for the Villain
   interaction between nodes.  For the cosine interaction however, the
corresponding
   transitions will occur at $lower$ temperatures than in the Villain case
   (due to the additional reduction in junction coupling resulting from the
   vortex-spinwave interaction which is present in the cosine model).
   Thus these transitions will be even more obscured by high energy barriers
   than our comparison might indicate. Using the same method we have also
computed the barriers for the Villain interaction and we have found that for
all types of networks these are even much higher (typically by one order
of magnitude) than those cited for the cosine model. For example,
we have computed $E_b^{Villain}=0.517$ for a square network, at low
temperature. This large difference comes predominantly from the bond that
is crossed by the vortex when climbing to the neighboring site, and it can
be easily understood by comparing Villain and cosine potentials.
 We note also that, although
   these barriers are important for the dynamics of real networks, they do not
   effect the equilibration of our simulation; this is because in our Coulomb
   gas MC, we  move vortices in discrete steps to neighboring cells, without
the need to climb over the energy barrier.  However had we done typical
Metropolis MC
   in terms of the phases $\theta_i$ and the Hamiltonian of Eq.(\ref{eqH}),
   these barriers would cause equilibration problems; this is because
   when a vortex moves from one cell to its neighbor, the $\theta_i$
   evolve along a continuous path in phase space, that therfore must
   take the system over the energy barrier separating the two cells.
\bibitem{dang} E. K. F. Dang and B. L. Gy\"orffy, \prb{\bf 47}, 3290 (1993).
\bibitem{expt33} H. S. J. van der Zant, H. A. Rijken, and J. E. Mooij,
    J. Low Temp. Phys. {\bf 82}, 67 (1991).
\bibitem{plasma} Ph. Choquard and J. Clerouin, \prl{\bf 50}, 2086 (1983);
	J. M. Caillol, D. Levesque, J. J. Weis, and J. P. Hansen, J. Stat. Phys.
	{\bf 28}, 325 (1982).
\bibitem{expand} E. Br\'{e}zin, A. Fujita and S. Hikami, \prl{\bf 65}, 1949
(1990);
 S. Hikami, A. Fujita and A. I. Larkin, \prb{\bf 44}, 10400 (1990).
\bibitem{oneill} J. A. O'Neill and M. A. Moore, \prl{\bf 69}, 2582 (1992);
	\prb{\bf 48}, 374 (1993).
\bibitem{LLL1} Z. Te\v{s}anovi\'{c} and L. Xing, \prl{\bf 67}, 2729 (1991).
\bibitem{LLL2}   Jun Hu and A. H.  Macdonald, \prl{\bf 71}, 432 (1993).
\bibitem{LLL3} Y. Kato and N. Nagaosa, \prb{\bf 47}, 2932 (1993); \prb {\bf
48}, 7383 (1993).
\bibitem{sasik} R. \v{S}\'{a}\v{s}ik and D. Stroud, \prb{\bf 49}, 16074 (1994).
\bibitem{tesanovic} I.. F. Herbut and Z. Te\v{s}anovi\'c, \prl{\bf 73}, 484
(1994).


\end{thebibliography}

%%%%%%%%%%%%%%%%%%%%%%%%%%%%%%%%%%%%%%%%%%%%%%%%%%%%%%%%%%%%%%%%%%%%%%%%%%
\newpage

\begin{table}
\begin{center}
\begin{minipage}{10cm}
\begin{tabular}{|c|c|c|c|}
  $T$  &  $\eta_{{\bf G}_1}(T) $ &  $\eta_{{\bf G}_2}(T) $ &
  $\varphi_6^\infty$   \\ \hline
 0.00475 & 0.188 $\pm$0.008   & 0.704 $\pm$0.055  & 0.571 $\pm$0.007  \\
 0.00500 & 0.207 $\pm$0.007   & 0.806 $\pm$0.032  & 0.529 $\pm$0.005  \\
 0.00525 & 0.211 $\pm$0.007   & 0.852 $\pm$0.028  & 0.504 $\pm$0.004  \\
 0.00550 & 0.248 $\pm$0.005   & 0.998 $\pm$0.019  & 0.476 $\pm$0.003  \\
 0.00575 & 0.255 $\pm$0.008   & 0.999 $\pm$0.029  & 0.458 $\pm$0.003  \\
 0.00600 & 0.296 $\pm$0.006   & 1.065 $\pm$0.028  & 0.426 $\pm$0.007  \\
 0.00625 & 0.319 $\pm$0.010   & 1.191 $\pm$0.016  & 0.403 $\pm$0.004  \\
 0.00650 & 0.4   $\pm$0.16    & 1.4   $\pm$0.22   & 0.33  $\pm$0.030  \\
 0.00675 & 1.4   $\pm$0.31    & 2.0   $\pm$0.31   & 0.20  $\pm$0.041  \\
 0.00750 & 3.4   $\pm$0.37    & 3.4   $\pm$0.44   & 0.03  $\pm$0.046  \\
 0.01100 & 2.8   $\pm$0.23    & 2.9   $\pm$0.30   &-0.01  $\pm$0.032  \\
 0.01500 & 2.2   $\pm$0.12    & 2.1   $\pm$0.22   & 0.00 $\pm$0.020   \\
\end{tabular}
\end{minipage}
\end{center}

\caption[]{ Temperature dependence of the exponents $\eta_{{\bf G}_1}(T)$ and
$\eta_{{\bf G}_2}(T)$ for $f=1/49$ on the triangular grid, as obtained from
finite size scaling. Also displayed are the limiting values $\phif$ of the
orientational correlation
$\varphi_6(T)$ for $L\rightarrow \infty$.}
\label{t1}
\end{table}

\begin{table}
\begin{center}
\begin{minipage}{11cm}
\begin{tabular}{|c|c|c|c|c|}
    $T$  & \multicolumn{4}{c|}{$\eta_{{\bf G}_1}(T)$}  \\ \cline{2-5}
         &  $L=63$ &  $L=77$ & $L=91$ &  FSS \\ \hline
 0.00475 & 0.164$\pm$0.026 & 0.161$\pm$0.025 & 0.195$\pm$0.018 & 0.188
$\pm$0.008 \\
 0.00500 & 0.216$\pm$0.012 & 0.221$\pm$0.009 & 0.219$\pm$0.007 & 0.207
$\pm$0.007 \\
 0.00525 & 0.249$\pm$0.009 & 0.235$\pm$0.009 & 0.247$\pm$0.006 & 0.211
$\pm$0.007 \\
 0.00550 & 0.282$\pm$0.007 & 0.272$\pm$0.006 & 0.275$\pm$0.007 & 0.248
$\pm$0.005 \\
 0.00575 & 0.298$\pm$0.008 & 0.292$\pm$0.009 & 0.290$\pm$0.006 & 0.255
$\pm$0.008 \\
 0.00600 & 0.326$\pm$0.008 & 0.326$\pm$0.007 & 0.329$\pm$0.007 & 0.296
$\pm$0.006 \\
 0.00625 & 0.351$\pm$0.014 & 0.350$\pm$0.017 & 0.352$\pm$0.016 & 0.319
$\pm$0.010 \\
 0.00650 & 0.677$\pm$0.342 & 0.473$\pm$0.223 & 0.568$\pm$0.131 & 0.4 $\pm$0.16
\\
 0.00675 & 1.755$\pm$0.789 & 1.678$\pm$0.453 & 2.458$\pm$0.911 & 1.4 $\pm$0.31
\\
\end{tabular}
\end{minipage}
\end{center}
\caption[]{Comparison of the exponents $\eta_{{\bf G}_1}(T)$ obtained using two
different methods. Coulumns 2-4 show results from fitting of $S(\bG)$ to
Eq.(\ref{e21.2}), for system sizes $L=63,77,91$. Coulumn 5, labelled FSS,
restates the results from the finite size scaling analysis. All exponents are
for density $f=1/49$. }
\label{t2}
\end{table}

\begin{table}
\begin{center}
\begin{minipage}{4.5cm}
\begin{tabular}{|c||c|}
  $T$  &  $\eta_{{\bf G}_1}(T) $ \\ \hline
 0.0040 & 0.0024$\pm$0.001 \\
 0.0045 & 0.111 $\pm$0.016 \\
 0.0050 & 0.198 $\pm$0.012 \\
 0.0055 & 0.224 $\pm$0.009 \\
 0.0060 & 0.270 $\pm$0.011 \\
 0.0065 & 0.33  $\pm$0.04  \\
 0.0070 & 0.49  $\pm$0.10  \\
\end{tabular}
\end{minipage}
\end{center}

\caption[]{ Temperature dependence of the exponents $\eta_{{\bf G}_1}(T)$
of the floating vortex lattice on the square grid as obtained by fitting the
height of peaks in the structure function, for $f=1/60$ and $L=60$.}
\label{t3}
\end{table}

%%%%%%%%%%%%%%%%%%%%%%%%%%%%%%%%%%%%%%%%%%%%%%%%%%%%%%%%%%%%%%%%%%%%%%%%%%%
\begin{figure}
\caption[]{ Phase diagram of the sufficiently dilute system, as found by our
Monte Carlo calculation.}
\label{f1}
\end{figure}

\begin{figure}
\caption[]{ Structure function $S({\bf k})$ in the first Brillouin zone,
(upper portion of the figure), and the profile of the peak heights along the
vertical symmetry axis $k_y$ (lower portion);
 for $f=1/49$ and $N_c=63$, and three different temperatures $T$:
(a) $T=0.003$, just below $T_c$, in the
``pinned solid" phase, (b) $T=0.0065$, just below $T_m$, in the
``floating solid" phase, (c) $T=0.0075$, just above $T_m$, in the liquid.
Intensities in the density plots are plotted nonlinearly to enhance features.}
\label{f2}
\end{figure}

\begin{figure}
\caption[] {Inverse dielectric function $\epsilon^{-1}(T)$ and
orientational order correlation $\varphi_6(T)$ versus $T$
for $f=1/49$ and $N_c=169$. Solid and dashed lines are guides to the eye only.}
\label{f3a}
\end{figure}

\begin{figure}
\caption[]{The  dependence of the depinning
and melting temperatures, $T_c$ and $T_m$, on vortex density $f$. Errors are
estimated from the width in the apparent drop in $\epsilon^{-1}(T)$ and
$\varphi_6(T)$. Solid and dashed lines are guides to the eye only.}
\label{f3b}
\end{figure}

\begin{figure}
\caption[]{Finite size scaling of $S({\bG_1})/L^2$ (note the log-log
scale) for $f=1/49$. Solid and dashed lines are fits to Eq.\ (\ref{e15}b). }
\label{f4a}
\end{figure}

\begin{figure}
\caption[]{ Heights of the peaks $S(\bG)$ versus $|\bG|$ for $f=1/49$ and
$L=63$.
Dashed lines represent the best fit to Eq.(\ref{e21.2}), and are used to
extract the exponent $\eta_{\bG_1}(T)$.}
\label{f6}
\end{figure}

\begin{figure}
\caption[]{
Finite size scaling of $\varphi_6(T)$ for $f=1/49$.
Solid and dashed lines are fits to Eq.\ (\ref{e16}a).}
\label{f4b}
\end{figure}

\begin{figure}
\caption[]{ $T/\eta_{\bG_1}(T)$ (proportional to the  shear modulus $\mu$)
 and orientational  correlation $\phif$ versus $T$, as extracted from
 finite size scaling. The intersection of $T/\eta_{\bG_1}(T)$ with the dashed
line $3T$ determines the KTNHY upper bound on the melting transition
  $T/\eta_{\bG_1}(T) > 3T$.}
\label{f5a}
\end{figure}

\begin{figure}
\caption[]{Comparison
of $\eta^{-1}_{\bG_1}(T)$ and $-1/K\ln[\phif(T)]$ as a test of Eq.(\ref{e14}).
 Below $T_m$ the two  quantities
coincide, as expected  for a 2D harmonic lattice. Dashed line determines the
KTNHY upper bound $\eta^{-1}_{\bG_1}(T) > 3$.}
\label{f5b}
\end{figure}

\begin{figure}
\caption[]{Translational and orientational correlation lengths $\xi_+(T)$ and
$\xi_6(T)$ versus $T$ for $f=1/49$, as extracted from  finite size scaling.
Both correlation lengths sharply increase at the melting temperature
$T_m\simeq 0.0070$.}
\label{f7}
\end{figure}

\begin{figure}
\caption[]{ Free energy distribution $F(E)$ versus $E$, at melting $T_m$,
for $f=1/49$ and several system sizes $L$.
 The growth in the energy barrier $\Delta F$ with
increasing $L$ (see inset) indicates a first order transition.  Curves
for different $L$ are offset from each other by a constant, for the sake
of clarity.}
\label{f8}
\end{figure}

\begin{figure}
\caption[]{ Free energy distribution $F(E)$ versus $E$, at the depinning
transition $T_c$, for $f=1/49$ and
 several system sizes $L$.  The growth in energy barrier $\Delta F$ with
increasing $L$ (see inset) indicates a first order transition.  Curves
for different $L$ are offset from each other by a constant, for the sake
of clarity. The abrupt ending of the distributions at the low energy side of
the
graph is because  the lower minimum represents the ground state energy.}
\label{f9}
\end{figure}

\begin{figure}
\caption[]{Dependence of $T_c$ and $T_m$ on vortex density $f$ for the dilute
system
on a square grid. Dashed and solid lines are guides to the eye only.}
\label{f10}
\end{figure}

\begin{figure}
\caption[]{Two types of  ground state configurations for a dilute system
on the square grid: (a) nearly triangular vortex lattice $f=1/60$; (b) nearly
square vortex lattice $f=1/51$. Solid squares denote positions of vortices.}
\label{f11}
\end{figure}

\begin{figure}
\caption[]{(a) Inverse dielectric constant $\epsilon^{-1}(T)$ and
(b) orientational correlations $\varphi_6(T)$ and $\varphi_4(T)$ versus $T$,
 for the system with nearly triangular ground state with $f=1/60$ and
$L=60$.}
\label{f12}
\end{figure}

\begin{figure}
\caption[]{Melting of a nearly triangular vortex lattice on the square grid.
Intensity plots of $S(\bk)$ for $f=1/60$, $L=60$ and
several temperatures: (a) $T=0.003$ in the pinned solid; (b) $T=0.006$
floating solid; (c) $T=0.009$ in the liquid.}
\label{f13}
\end{figure}

\begin{figure}
\caption[]{ Heights of the peaks $S(\bG)$ versus $|\bG|$ for $f=1/60$ and
$L=60$.
Dashed lines represent the best fit to Eq.(\ref{e21.2}), and are used to
extract the exponent $\eta_{\bG_1}(T)$.}
\label{f14}
\end{figure}

\begin{figure}
\caption[]{(a) Inverse dielectric constant $\epsilon^{-1}(T)$ and
(b) orientational correlations $\varphi_6(T)$ and $\varphi_4(T)$ versus  $T$,
for the system with nearly square ground state with $f=1/51$ and $L=51$.}
\label{f15}
\end{figure}

\begin{figure}
\caption[]{Melting of a nearly square vortex lattice on the square grid.
Intensity plots of $S(\bk)$ for $f=1/51$, $L=51$ and
several temperatures: (a) $T=0.003$ in the pinned solid; (b) $T=0.0045$
in the floating
solid; (c) $T=0.006$ in the liquid.}
\label{f16}
\end{figure}

\begin{figure}
\caption[]{Ground state for $f=5/11$ on a  $22\times 22$ unit cell. Solid
squares
represent vortex positions. Crosses $(+)$
indicate vacancies (defects) in the otherwise perfect checkerboard pattern
of $f=1/2$.}
\label{f17}
\end{figure}

\begin{figure}
\caption[]{Melting of $f=5/11$ for $L= 22$.  $S(\bk)$  is
shown for: (a) $T=0.010$ in the pinned defect solid; (b) $T=0.018$ in the
defect liquid; (c) $T=0.055$ in the completely disordered high temperature
phase.}
\label{f18}
\end{figure}

\begin{figure}
\caption[]{Peak heights $S(\bq)^*$ with $\bq^*\equiv(\pi/a_0,\pi/a_0)$, and
$S(\bG_1)$
where $\bG_1=(2\pi/L)(1,5)$ is the shortest reciprocal lattice vector of the
defect superlattice, plotted versus $T$
 for $f=5/11$ and $L=22$. }
\label{f19}
\end{figure}

\begin{figure}
\caption[]{Inverse dielectric constant $\epsilon^{-1}(T)$ versus $T$
for $f=5/11$ and $L=22$.}
\label{f20}
\end{figure}

\begin{figure}
\caption[]{Melting of the defect superlattice for $f=22/45$ and $L=90$.
Intensity plot of $S(\bk)$ for: (a) $T=0.0040$ in the pinned defect
solid; (b) $T=0.0065$ in the defect floating solid; (c) $T=0.0085$ in the
defect
liquid. Intensities at $(\pm\pi/a_0,\pm\pi/a_0)$ and $(0,0)$, that arise
from
the fixed staggered background, have been substracted for the sake of clarity.
}
\label{f21}
\end{figure}

\begin{figure}
\caption[]{Inverse dielectric function $\epsilon^{-1}(T)$ and orientational
correlation $\varphi_6(T)$ versus $T$, for $f=22/45$ and $L=90$. }
\label{f22}
\end{figure}

%\begin{figure}
%\caption[]{}
%\label{f}
%\end{figure}

\end{document}